\documentclass[numberedappendix]{emulateapj}
\usepackage{epsfig}
\usepackage{graphicx}
\usepackage{color}
\usepackage{enumerate}
\usepackage{ulem}

\def\beq{\begin{equation}}
\def\eeq{\end{equation}}

\def\pc{{\rm\,pc}}

\def\yr{{\rm\,yr}}
\def\Myr{{\rm\,Myr}}
\def\Gyr{{\rm\,Gyr}}

\def\kms{{\rm\,km\,s^{-1}}}
\def\kpc{{\rm\,kpc}}

\def\mbh{M_{\rm BH}}
\def\rbh{r_{\rm BH}}
\def\Mdot{\dot{M}}
\def\rt{r_{\rm t}}

\def\rsch{r_{\rm sch}}

\def\msun{\rm M_\odot}

\def\mstar{\rm m_*}
\def\rstar{\rm r_*}
\def\acc{\rm M_\odot~yr^{-1}}

\def\nbody{$N$-body~}
\def\rinf{r_{\rm h}}
\def\Mbh{M_{\rm BH}}
\def\thetaD{\theta_{\rm D}}
\def\thetatwo{\theta_{\rm 2}}
\def\thetalc{\theta_{\rm{lc}}}
\def\rcrit{r_{\rm {crit}}}

\newcommand{\accrate}{{\rm M_\odot~yr^{-1}}}

\shorttitle{TIDAL DISRUPTION RATE EVOLUTION FOR SMBHB}
\shortauthors{Li et al.}

\begin{document}

\title{BOOSTED TIDAL DISRUPTION BY MASSIVE BLACK HOLE BINARIES DURING GALAXY MERGERS FROM THE VIEW OF $N$-BODY SIMULATION}

\author{Shuo Li\altaffilmark{1,2}, F.K. Liu\altaffilmark{2,3}, Peter Berczik\altaffilmark{1,4,5}, \& Rainer Spurzem\altaffilmark{1,4,3}}

\altaffiltext{1}{National Astronomical Observatories and Key Laboratory of Computational Astrophysics, Chinese Academy of Sciences, 20A Datun Rd., Chaoyang District, Beijing 100012, China; lishuo@nao.cas.cn}
\altaffiltext{2}{Department of Astronomy, School of Physics, Peking University, Yiheyuan Lu 5, Haidian Qu, Beijing 100871, China}
\altaffiltext{3}{Kavli Institute for Astronomy and Astrophysics, Peking University, Yiheyuan Lu 5, Haidian Qu, Beijing 100871, China}
\altaffiltext{4}{Astronomisches Rechen-Institut, Zentrum f\"{u}r Astronomie, University of Heidelberg, M\"{o}chhofstrasse 12-14, Heidelberg 69120, Germany}
\altaffiltext{5}{Main Astronomical Observatory, National Academy of Sciences of Ukraine, 27 Akademika Zabolotnoho St., 03680 Kyiv, Ukraine}

\begin{abstract}

Supermassive black hole binaries (SMBHBs) are productions of the hierarchical galaxy formation model. There are many close connections between central SMBH and its host galaxy because the former plays very important roles on the formation and evolution of a galaxy. For this reason, the evolution of SMBHBs in merging galaxies is an essential problem. Since there are many discussions about SMBHB evolution in gas rich environment, we focus on the quiescent galaxy, using tidal disruption as a diagnostic tool. Our study is based on a series of numerical large particle number direct \nbody simulations for dry major mergers. According to the simulation results, the evolution can be divided into three phases. In phase I, the tidal disruption rate for two well separated SMBHs in merging system has similar level to single SMBH in isolate galaxy. After two SMBHs getting close enough to form a bound binary in phase II, the disruption rate can be enhanced for $\sim 2$ order of magnitudes within a short time. This "boosted" disruption stage finishes after the SMBHB evolving to compact binary system in phase III, corresponding to a drop back of disruption rate to a level of a few times higher than that in Phase I. How to correctly extrapolate our \nbody simulation results to reality, and implications of our results to observations, are discussed too.

\end{abstract}


\section{Introduction}
\label{intro}

In the $\Lambda$ cold dark matter ($\Lambda$CDM) cosmology, supermassive black hole (SMBH) binaries (SMBHBs) are the direct descendent of hierarchical galaxy mergers \citep{bege80,volo03}. If two merging galaxies are gas-rich with comparable masses, a large fraction of gas may be driven into galactic center, prompting a burst of star formation and feeding the central SMBH \citep{barn91,miho94,spri05a,hopk05}. As a result, an active galactic nucleus (AGN) form in the center of galaxy and the feedback provoked by accreting SMBH may regulate the mass growth both of the central black hole and galaxy \citep{spri05b,sija07,matt08,boot09}, which might be the physics mechanisms forming the observed tight correlations of the central SMBH and host galaxy \citep{mago98,ferr00,gebh00,trem02,korm13}. Attribute to the progress on cosmological simulation in recent years, more details about black hole growth and SMBH-host galaxy co-evolution have been revealed \citep{hirs12,choi14,sija15,kann15}. Recent investigations suggested that minor or dry mergers without AGN activities play important roles in the formation and evolution of both late- and massive early- types of galaxies \citep{naab09,dokk10}. Quiescent SMBHB may form in all the kinds of galaxy mergers.

To become gravitationally bound and finally coalesced, two SMBHs in merging galaxies have to lose effectively their large angular momentum and decrease the separation. It has been shown that the dynamical evolution of two SMBHs could be divided into several stages \citep{bege80}. At the beginning of merger, two SMBHs with the galactic cores inspiral toward each other because of galactic dynamical friction \citep{chan43}. As two SMBHs become closer and closer, more and more ambient stars around SMBHs are stripped away by background potential. When the separation of two SMBHs is about their influence radius, the two SMBHs become gravitationally bound and a SMBHB forms. During the further evolution of SMBHB, more and more stars in the nuclear cores are scattered off the system through three-body slingshot effect and the dynamical friction becomes less and less efficient in hardening the SMBHB. When most of the stars bound to the SMBHs are ejected away, the dynamical friction becomes inefficient. Then the evolution of SMBHB is dominated by the slingshot effect of SMBHB scattering off stars and depends on the replenishment of the scattered stars \citep{sasl74,mikk92,quin96}. If the replenishment of the scattered stars is dominated by the two-body relaxation process in a system of spherically symmetric distributions, the slingshot effect will be inefficient and the SMBHB may stall at parsec scale for a time-scale longer than the Hubble time, which is the so-called "final parsec problem" in SMBHB dynamical evolution \citep{bege80,milo01,ber05}. However, both analytic calculations and N-body simulations in the past decade suggested that the "final parsec problem" can be overcome in realistic systems of galaxy mergers due either to gas dynamics or to stellar dynamics other than spherical two-body relaxation \citep[][and references therein]{goul00,yu02,mer04,ber06,pret11,khan11,khan13,colp14}. If an SMBHB can be driven to a separation of milliparsec scale, the strong gravitational wave (GW) radiations would efficiently remove its orbital energy and angular momentum, and drive it coalesced within a Hubble time \citep{pet64,bege80}. \nbody numerical simulation results recently indicated that, the coalescence timescales for this stage can be less than 1\Gyr, or even much shorter for SMBHBs with relatively high eccentricity orbits \citep{khan12,sobo15}. That also has been confirmed under more general conditions, such as non-spherical galaxies and the collisionless limit using a novel Monte Carlo method instead of \nbody \citep{vas15}, and in scattering experiments \citep{ses15}. Due to the anisotropy of GW emissions during coalescence, the remnant of the post-coalesced SMBHs can be kicked with a recoil velocity from a few hundreds of kilometers per second as usual up to several thousands of kilometers per second in extreme cases\citep{pere62,bake06b,kopp07,camp07,lous11}. The strong GW radiations by SMBHBs are the main targets of the planed evolved {\it Laser Interferometer Space Antenna} (e{\it LISA}) and the ongoing GW detection program Pulsar Timing Array (PTA)\citep{cons13,ver16}.

A number of observational evidences for SMBHBs in gas rich environments were reported in the literature, ranging from resolved binary AGNs \citep{komo03,huds06,rodr06,fabb11,dean14} and candidates for unresolved binary systems in AGNs with double-peaked broad lines \citep{tsal11,ju13,shen13,liu14b}, helical radio jets \citep{bege80,conw95,liu07,rola08,kun14}, quasi-periodic variabilities \citep[][and references therein]{sill88,liu95,liu02,liu06,valt11,grah15}, and X-shaped radio structures \citep{liu04}, to candidates for the coalescence remnants of SMBHs in AGNs of interruption and recurrence of jet formation in double-double radio galaxies \citep{liu03} and of characteristic signatures of recoiling SMBHs \citep{komo08b,liu12,civa12,lena14}. However, due to the native "black" property, SMBHBs in gas poor environments are extremely hard to detect. A quiescent SMBH can be temporarily illuminated and investigated by tidally disrupting stars passing by \citep{hil75,ree88,evan89,guil13}, causing giant flares from $\gamma$-ray to radio bands \citep[][and references therein]{geza13,komo15}. It was suggested first theoretically and then confirmed in X-ray observations that a quiescent SMBHB in galaxy can also be probed with the observations of tidal disruption events (TDEs). Based on the theoretical investigations, \citet{liu09} showed that when a TDE occurs in a SMBHB system, the gravitational perturbation of the companion SMBH would cause characteristic drops in the light curve of TDE. The characteristic drops was observed recently in the X-ray light curve of the TDE candidate SDSS J120136.02+300305.5 and consistent exclusively with the predication of the binary model \citep{liu14a}.

In the past decade, about 30-40 TDEs have been detected. According to analytical and numerical estimations, the rate of TDEs by single SMBH systems in isolated spherical galaxies is $\sim 10^{-6}-10^{-4}$~$\acc$ per galaxy or a few times higher in flattened systems \citep{syer99,mago99,wan04,broc11,vasi13,velz14,zho14,ston16}. It also can be impacted if we take into account tidal disruption for giant stars \citep{macl12}. The rate of TDEs by gravitationally recoiling single SMBHs is several orders of magnitude smaller \citep{komo08,olea12,ston11,li12,ston12}. Besides, the TDE rate of SMBHBs in galaxy mergers was investigated analytically and/or with scattering experiments. \citet{chen08} showed that the TDE rate of hard SMBHB systems is orders of magnitude lower than that in single SMBH systems if the SMBHB is in a spherical isotropic galactic cores and two-body relaxation is dominated. However, for a real SMBHB system, two SMBHs are bounded by nuclear star clusters. Both analytic calculations and numerical scattering experiments indicated that the perturbations of the companion SMBH in a self-gravity bound SMBHB system would scatter a large fraction of the bound stars toward the central SMBH and significantly enhance the TDE rate by several orders of magnitude up to one event per galaxy per year \citep{ivan05,chen09,chen11,wegg11}. However, in their calculations, the contributions of the stars bound to the companion SMBH and the un-bound/weakly bound stars due to two-body relaxation and/or triaxial gravitational potential did not be taken into account. The analytic investigations showed that the TDE rate of stars by SMBHs in the early phases of galaxy mergers when galactic dynamical friction is dominated could also be enhanced by several orders of magnitude up to $10^{-2}$ events per year per galaxy because the perturbation of companion galactic core and the triaxial distribution of gravitational potential of the galactic nucleus \citep{liu13}.

Because the limitations of analytic calculations and scattering experiments, the investigations of TDE rates in literature have to be done by splitting the evolution of SMBHBs in galaxy mergers into several stages based on the dominant physical processes. Other important issues which are not always properly taken into account are two-body relaxation between the stars around the SMBHB, the contribution of stars to the gravitational potential, and the evolution of the SMBHB with regard to its orbital parameters. Our current paper aims to resolve these issues. In this paper, we investigate the dynamical evolution of SMBHBs in major galaxy mergers with direct N-body simulations and self-consistently calculate the TDE rates by SMBHBs from the early phase of galaxy mergers to the post-merger of galaxies. A practical difficulty is that we are not able yet to accurately simulate a full galaxy with bulge and nucleus star-to-star due to computational limitations. An interesting attempt is to use expansion techniques on the Poisson equation to speed up the integration, which has been announced by \citet{meir14} recently. However, to use that method in our model, there are several modifications have not been finished yet. As a result, we use the method of scaling from smaller particle number $N$ and larger tidal radius to realistic value. Our results show that the TDE rates by SMBHBs in major galaxy mergers can be enhanced by up to two orders of magnitude, which is consistent with the high detection rates of TDEs preferred in E+A galaxies \citep{arca14}. That is very interesting because E+A galaxies are believed to be the post-mergers galaxies \citep{zab96,got05,ston16}.

This paper is organized as follow. In Section~\ref{theory}, we briefly introduce the loss cone theory for galaxy mergers. In Section~\ref{mtd}, we describe our \nbody simulation method and the galaxy model. The numerical results of TDE rate and its variations as function of the dynamical evolution of SMBHBs in galaxy mergers are given in Section~\ref{res}. We discuss and extrapolate our simulation results to a few typical realistic systems of galaxy mergers in Section~\ref{Extra}, and have a short discussion about implications to observations in Section~\ref{obv}. Section~\ref{sum} gives a short summary of our results.

\section{Tidal Disruption in Merging System}
\label{theory}

Due to strong tidal force from black hole (BH), a star will be tidally disrupted if it can penetrate into the vicinity of a BH within $r<\rt$ \citep{hil75,ree88}
\beq
\rt\backsimeq \mu\rstar(\Mbh/\mstar)^{1/3}.
\label{eq:rt}
\eeq
Here $\mu$ is a dimensionless parameter of order unity. $\rstar$, $\mstar$ and $\Mbh$ are stellar radius, stellar mass and BH mass, respectively. Since $\rt\propto\Mbh^{1/3}$ while Schwarzschild radius $\rsch\propto\Mbh$, a star will be directly swallowed instead of tidally disrupted by a Schwarzschild SMBH with mass $\Mbh\gtrsim 10^8\msun$.

In a two body system with only one single BH and a star, if the star has specific angular momentum smaller than a critical tidal disruption (TD) angular momentum $J_{\rm{lc}}$, which satisfy $J\lesssim J_{\rm {lc}}\simeq(2G\Mbh\rt)^{1/2}$ , it will be tidally disrupted by the BH
within one orbit period. Such kind of stars always have their velocity vectors inside a so-called "loss cone" with opening angle \citep[][and references therein]{lig77, merr13}
\beq
\thetalc = \frac{J_{\rm lc}}{J_{\rm c}},
\label{eq:theta_lc}
\eeq
where $J_{\rm c}$ is the specific circular angular momentum with same energy of the star. In a spherical isotropic stellar distribution, for stars inside the influence radius $\rinf$ of SMBH, $r<\rinf$,
\beq
\theta^{2}_{\rm{lc}} \simeq \frac{2}{3}\frac{r_{\rm t}}{r},
\label{eq:theta_lc2a}
\eeq
while for stars at $r>\rinf$,
\beq
\theta^{2}_{\rm{lc}} \simeq \frac{2}{3}\frac{r_{\rm t}\rinf}{r^2}.
\label{eq:theta_lc2b}
\eeq
\citep{fra76,bau04b}.

Due to the interaction with environment, the angular momentum of a star changes with time. If $\thetaD(r)$ is the orbital averaged deflection angle of velocity and $\thetaD(r)\ll\thetalc(r)$, all of the stars inside loss cone are in the diffusive regime and would be tidally disrupted within one orbital period \citep{fra76,shap76,lig77}. The loss cone becomes empty because its replenishment through diffusion takes time much longer than the stellar orbit periods. While, for pinhole regime with $\thetaD(r)\gg\thetalc(r)$, the loss cone keeps full because the star can be deflected into and out within one orbital period. Thus there is a critical radius $\rcrit$, which gives $\thetaD(r_{crit})\sim\thetalc(r_{crit})$. Detailed theoretical calculations and numerical simulations show that the TD rate of stars peaks at about $\rcrit$ \citep{fra76,lig77,amar04,zho14}.

In a spherical system, the disruption rate contributed by the stars between $r$ and $r+\rm dr$ is
\beq
d\Gamma \backsimeq \frac{4\pi r^2 \rm dr\rho(r)}{m_*}\frac{\theta^2(r)}{t_{\rm d}(r)}
\label{eq:TDNRT}
\eeq
\citep{fra76,syer99,liu13}, where $t_{\rm d}(r)$ is dynamical timescale of the star, $\rho$ is stellar mass density, and $\theta$ is a dimensionless coefficient for stars depleting into the loss cone. Detailed analyze \citep{you77} suggests that
\beq
\theta^2=\min(\thetalc^2, \thetaD^2/\ln\thetalc^{-1}) .
\label{eq:thetasq}
\eeq
To investigate the TD rate in a merging system, we adopt the same scheme as \citet{liu13} did, which writes the effective deflection angle as
\beq
\thetaD^2 = \theta_{\rm 2}^2 + \theta_{\rm p}^2 + \theta_{\rm c}^2 ,
\label{eq:theta_d2}
\eeq
where $\theta_{\rm 2}$, $\theta_{\rm p}$ and $\theta_{\rm c}$ are, respectively, the contribution from two-body relaxation, massive perturber and triaxial gravitational potential. Based on our \nbody simulation results presented in Section~\ref{res}, we can analyze which mechanism dominates the TDE rate in different stages, and make analytical estimations. As we will present in Section~\ref{Extra}, $\theta_{\rm 2}$ and $\theta_{\rm p}$ can be well estimated by Equation~(\ref{eq:thetatwo_2}) and ~(\ref{eq:theta_p2}), respectively. While $\theta_{\rm c}$ is too complicate to be estimated. Detailed discussions can be found in Section~\ref{Extra}.

The total disruption rate can be obtained by integrating Equation~(\ref{eq:TDNRT}). That means we need to know $\rho$, $t_{\rm d}$, and $\theta$, which are time-dependent in merging galaxies and evolving SMBHBs. For this reason, here we address the questions with direct $N$-body simulations.

\section{Direct N-body simulations of the TDE rates by SMBHs in merging galaxies}
\label{mtd}

The dynamical evolution of two SMBHs in merging galaxies has been investigated intensively with direct \nbody simulations in the literature \citep{cha03,ber06,pret11,khan11,spu12,gua12,khan13,wan14,vas14}, but none of them has taken into account the TD of stars by SMBHs. Here we use the same scheme in the direct \nbody simulations to calculate the TDE effects as was used by \citet{li12}. We give a brief description of the scheme and the interesting readers are refereed to that paper. A particle is considered to be tidally disrupted and removed from the system, when its distance to one of the SMBH less than a fixed tidal radius $\rt$. The total mass and linear momentum of the disrupted star is transferred to the SMBH. We neglect the growth of tidal radius because the tidal radius $\rt$ is relatively insensitive to the change of the SMBH mass $\mbh$. Since the mass of disrupted stars will be added to SMBHs and all of stars have equal mass, the mass accretion rate of SMBH reflects the TDE rate.

In spite of powerful calculation capability from our special many-core hardware (\emph{laohu} GPU cluster), it is still prohibitive to carry out a direct \nbody simulation with sufficient particle numbers to resolve every star in a real galaxy. To circumvent this problem, we need to find out how the most important physical processes in the system depend on dimensionless parameters of the simulation such as the particle number $N$ and the adopted value for the tidal disruption radius $\rt$ relative to our simulation unit. Note that in our case we keep the ratio of the SMBH mass to the total mass of the system constant, motivated by observed correlations, so the number of stars $N$ is a proxy for the ratio between the single stellar mass and the SMBH mass. The goal is to provide several models in this two dimensional framework of $N$ and $\rt$, and then extrapolate for realistic values of both parameters.

Several problems need to be discussed when using this procedure. It is quite common in physics, but for our self-gravitating systems with huge dynamical range in spatial and time scales it needs some special attention. First, it is important that for every particle number $N$ the range of $\rt$ needs to be adjusted such that we get an appropriate loss cone - empty near the SMBH, going to full loss cones near a critical radius in the vicinity of the gravitational influence radius of the SMBH (see e.g. \citet{fra76} for the basis of this theory). If we would do a simulation with small $N$ and realistic (i.e. extremely small) $\rt$, all loss cones would be extremely small and full due to the short relaxation time, which is not the physical case. Therefore, for our range of particle numbers, only certain values of $\rt$ are physically relevant(see below), which make the critical radius large enough to have the empty to full loss cone transition at the right location. Second, the growth of the black hole per dynamical time becomes smaller and smaller for larger $N$ due to the increasing relaxation time. So, one may argue that our (relatively) small $N$ simulations contain some artificial mass growth of the SMBH, which in turn affects the potential and all the dynamics in the central region. The alternative, to neglect the SMBH mass growth artificially, however, creates more severe difficulties; it leads to spurious mass relocations (e.g. the SMBHB kicks out stars by slingshot which would otherwise be tidally accreted and disrupted, or if the stars are taken away the central gravitational potential is reduced artificially). We think that the best way to use the method of scaling is to include all physical processes (such as mass growth of the SMBHs) for every choice of parameters, and then carefully analyse the scaling behaviour with N and $\rt$ (including also e.g. the mass growth in the scaling analysis). We will come back to this issue in Section~\ref{Extra}.

In conclusion, we need to carry out a series of simulations with different $\rt$ and $N$ and then extrapolate the results to real systems. Similar strategy has been adopted in the literatures \citep[][and references therein]{bau04b,broc11,li12,zho14}.

In this paper we are focusing on a major merging system with two equal mass nuclei, which has spherical stellar distribution with equal mass stars respectively. For simplicity we choose a spherical Dehnen model to represent the nucleus for each galaxy \citep{deh93}. Within this model, the space density profile can be wrote as
\beq \rho(r)=\frac{3-\gamma}{4\pi}\frac{Ma}{r^\gamma(r+a)^{4-\gamma}} ,
\label{Dehnen rho}
\eeq
where $a$ is scaling radius, $M$ is total mass of each nucleus, and index $\gamma$ is a constant between interval [0,3). Since there are two galaxies with equal mass, the total mass for the merging system is $M_{\rm tot} = 2M$. And we can derive the cumulative mass for each nucleus by
\beq
M(r)=M\left(\frac{r}{r+a}\right)^{3-\gamma}.
\label{eq:mass}
\eeq
Thus the influence radius $\rinf$ which has been defined as $M_\star(r\leqslant \rinf)=2\Mbh$ is
\beq
\rinf=\frac{a}{(\frac{M}{2\Mbh})^{\frac{1}{3-\gamma}}-1},
\label{eq:rinf}
\eeq
and since $M\gg 2\Mbh$, $\rinf\approx a(2\Mbh/M)^{\frac{1}{3-\gamma}}$.

For simplicity, we adopt model units with $G=M=a=1$ thereafter, which is the same as \citet{li12}. Thus we have the same relations between our simulation units and physical quantities
\begin{eqnarray}
[T] &=& \left(\frac{GM}{a^3}\right)^{-1/2} \nonumber \\
    &=& 1.491\times 10^6(2^{\frac{1}{3-\gamma}}-1)^{3/2} \nonumber \\
      &\times& \left( \frac{M}{10^{11}\msun}\right)^{-1/2}\left(\frac{r_{1/2}}{1\kpc}\right)^{3/2} \yr, \\
    \label{eq:scalingT}
[V] &=& \left(\frac{GM}{a}\right)^{1/2} \nonumber \\
    &=& 655.8\times (2^{\frac{1}{3-\gamma}}-1)^{-1/2}\nonumber \\
    &\times& \left( \frac{M}{10^{11}\msun}\right)^{1/2}\left(\frac{r_{1/2}}{1\kpc}\right)^{-1/2} \kms, \\
    \label{eq:scalingV}
[R] &=& a=(2^{\frac{1}{3-\gamma}}-1)\left(\frac{r_{1/2}}{1\kpc}\right) \kpc, \\
    \label{eq:scalingr}
[\dot{M}] &=& M / [T] \nonumber \\
          &=& 6.707\times 10^4(2^{\frac{1}{3-\gamma}}-1)^{-3/2} \nonumber \\
          &\times& \left( \frac{M}{10^{11}\msun}\right)^{3/2}\left(\frac{r_{1/2}}{1\kpc}\right)^{-3/2} \msun/\yr.
          \label{eq:scalingMdot}
\label{eq:scalingr1}
\end{eqnarray}

Our fiducial galaxy model has been chosen as $\gamma = 1.0$, $M = 4\times 10^{10} \msun$, $\Mbh = 4\times 10^{7} \msun$, and half mass radius $r_{1/2} = 1 \kpc$. Thus we can derive $[T] \sim 0.63\Myr$, $[V] \sim 644\kms$ and $[R] \sim 0.4\kpc$.  For a solar type star, the tidal radius is $\sim 10^{-5} \pc$ corresponding to $\sim 2\times 10^{-8}$ in this simulation unit. Obviously, comparing to $\rt\sim 10^{-4}$ we have adopted in our \nbody simulations, this value is too small to have enough TDE records for statistical study in integration.

Before the integrations for merging system, each nucleus with central SMBH has been dynamically relaxed through the same method used in \citet{li12}. Similar to \citet{pret11} have adopted, here we initially set two nuclei with distance $d \sim20$, and a parabolic orbit with pericenter $\sim 1$. To accurately integrate the orbits of particles, we use a parallel direct \nbody $\varphi$\,-{\sc Grape}/$\varphi$\,-GPU code with fourth-order Hermite integrator, which is the same code adopted by \citet{li12}. For all of our integrations, the softening length has been set to $\epsilon=10^{-5}$, and the integrations have been terminated at $t=200$. All of the integrations are calculated on the $laohu$ GPU cluster in National Astronomical Observatories of China (NAOC).

In order to investigate the dependency of our results on particle number $N$, numerical tidal radius $r_{\rm t}$, density slope $\gamma$, and initial mass of SMBH, we have run a series of simulations with model parameters listed in Table~\ref{tab:para}. In Table~\ref{tab:para}, the first column is the sequence number of different simulations. Columns~2 - 6 are, respectively, initial orbital eccentricity $e_{\rm ini}$ of two galactic nuclei, particle number $N$ for each nuclei, numerical tidal radius $r_{\rm t}$, initial stellar density slope $\gamma$, and initial mass of SMBH. For comparison of the TDE rates, we have run four additional simulations for isolated single nucleus in model B01-B04. In Model C01, we have a tested a simulation for merging galaxy with moderate initial orbital eccentricity $e_{\rm ini} = 0.3$. All of simulations except the Model C01 are terminated at $t = 200$, when the separation of two SMBHs $r_{\rm BH}$ has $\rbh \ll \rinf$. The simulation of Model C01 is terminated at $t = 500$ because the dynamical evolution of the SMBHB is very slow for moderate initial eccentric orbit. In the simulation, Model A05 has been considered as our fiducial numerical model.

\begin{deluxetable*}{cccccc}
    \tablewidth{0pt}
    \tabletypesize{\scriptsize}
    \tablecaption{Parameters of simulation models \label{tab:para}}
    \tablehead{
    \colhead{$Model \, No.$} &
    \colhead{$e_{\rm {ini}}$} &
    \colhead{$N/Nucleus$} &
    \colhead{$r_{\rm t}$} &
    \colhead{$\gamma$} &
    \colhead{$\Mbh/M$} \\
    \colhead{(1)} &
    \colhead{(2)} &
    \colhead{(3)} &
    \colhead{(4)} &
    \colhead{(5)} &
    \colhead{(6)}
    }

    \startdata

A01 & 1.0  & $125\rm K$ & $5\times 10^{-4}$ & 1.0 & 0.001  \\
A02 & 1.0  & $250\rm K$ & $5\times 10^{-4}$ & 1.0 & 0.001  \\
A03 & 1.0  & $500\rm K$ & $5\times 10^{-5}$ & 1.0 & 0.001  \\
A04 & 1.0  & $500\rm K$ & $1\times 10^{-4}$ & 1.0 & 0.001  \\
A05 & 1.0  & $500\rm K$ & $5\times 10^{-4}$ & 1.0 & 0.001  \\
A06 & 1.0  & $500\rm K$ & $1\times 10^{-3}$ & 1.0 & 0.001  \\
A07 & 1.0  & $500\rm K$ & $5\times 10^{-4}$ & 0.5 & 0.001  \\
A08 & 1.0  & $500\rm K$ & $5\times 10^{-4}$ & 1.5 & 0.001  \\
A09 & 1.0  & $500\rm K$ & $5\times 10^{-4}$ & 1.0 & 0.0001  \\
A10 & 1.0  & $500\rm K$ & $5\times 10^{-4}$ & 1.0 & 0.01  \\
A11 & 1.0  & $1\rm M$ & $5\times 10^{-4}$ & 1.0 & 0.001  \\
\\
\hline \\
B01 & -isolated galaxy & $500\rm K$ & $5\times 10^{-5}$ & 1.0 & 0.001  \\
B02 & -isolated galaxy & $500\rm K$ & $1\times 10^{-4}$ & 1.0 & 0.001  \\
B03 & -isolated galaxy & $500\rm K$ & $5\times 10^{-4}$ & 1.0 & 0.001  \\
B04 & -isolated galaxy & $500\rm K$ & $1\times 10^{-3}$ & 1.0 & 0.001  \\
\\
\hline \\
C01 & 0.3  & $500\rm K$ & $5\times 10^{-4}$ & 1.0 & 0.001  \\

    \enddata

\tablecomments{ Col.(1): Model sequence number. Col.(2): Eccentricity of two nuclei at $t=0$. Col.(3): Particle numbers adopted in calculations for each nucleus. Col.(4): TD
radius $r_{\rm t}$. Col.(5): Density slope $\gamma$. Col.(6): Initial BH mass. All the integrations keep $\epsilon=10^{-5}$. For reference, we have four additional integrations in model B01-B04, with single nucleus and SMBH. And there is a model C01 with moderate initial eccentricity.}
\end{deluxetable*}

\section{Results}
\label{res}

\subsection{Evolution of tidal disruption}
\label{Evo_TD}

\begin{figure*}
\includegraphics[width=3.2in,angle=0.]{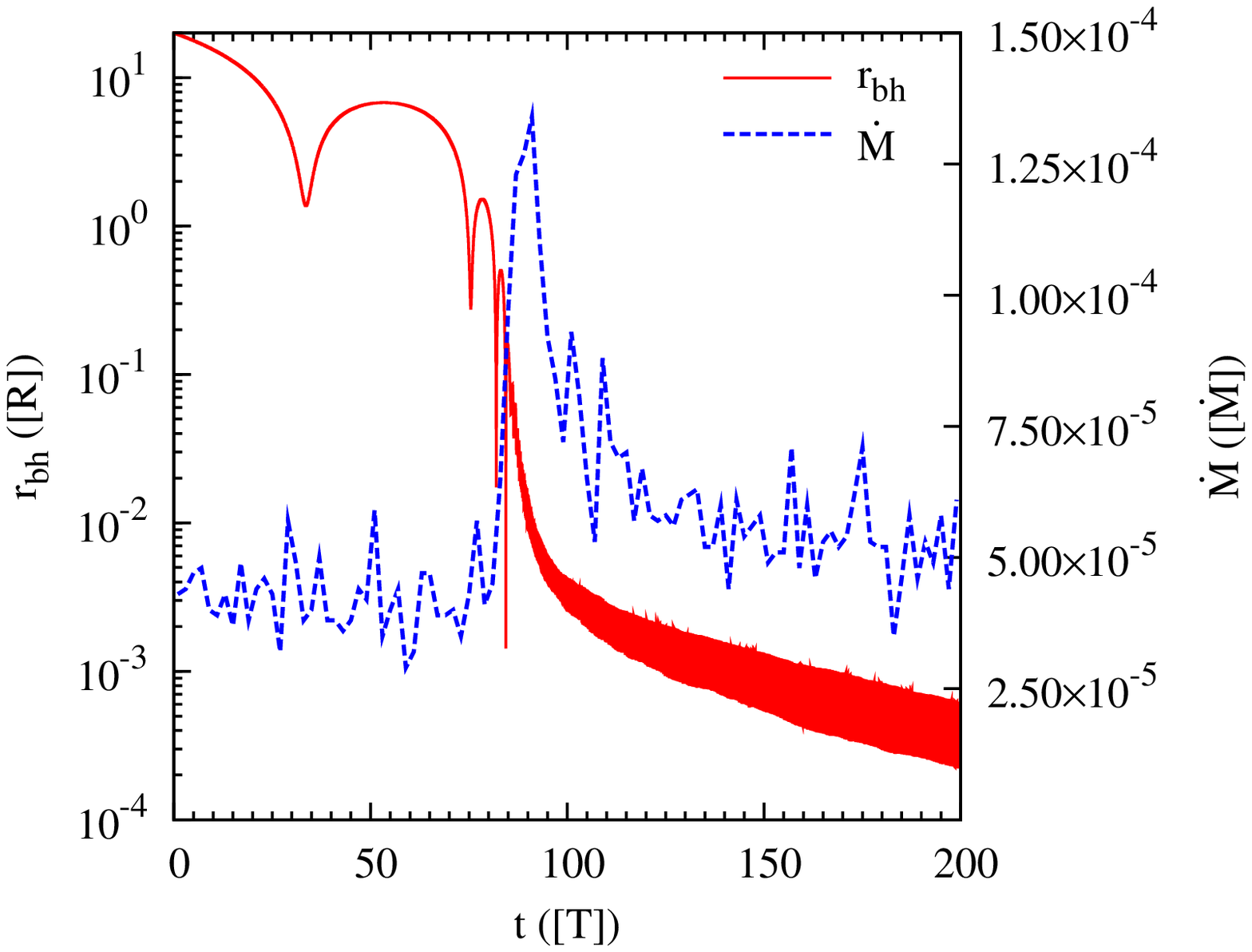}%
\hfill%
\includegraphics[width=3.2in,angle=0.]{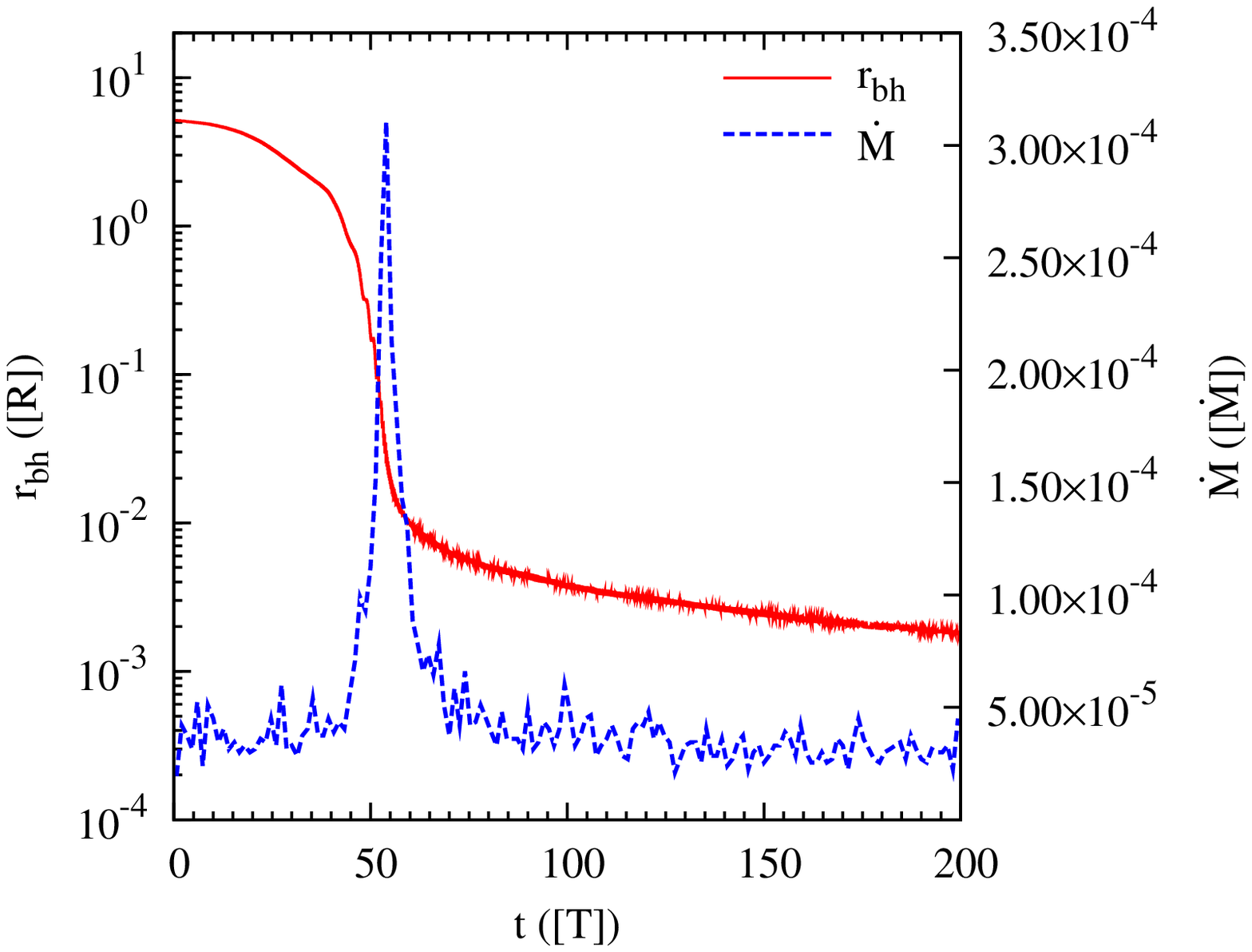}
\caption{Evolution of the separation for two SMBH and TD accretion rates. Here the \textit{x}-axis and the left \textit{y}-axis give evolved time and $\rbh$, the separation of SMBHs, respectively. The right \textit{y}-axis gives corresponding TD accretion rates $\Mdot$. Red solid line and blue dashed line represent $\rbh$ and $\Mdot$ respectively. Left panel: our fiducial model A05 with initial parabolic orbit. Right panel: model C01 with initial orbital eccentricity $e = 0.3$ for comparison. In order to compare to A05, here we only pick up the evolution for the last 200 time units. The $\Mdot$ in model C01 roughly keeps a constant level in the first 300 time units. Color version can be found in the electronic version \label{fig:TDrate}}
\end{figure*}

Fig.~\ref{fig:TDrate} gives the evolution of SMBHB orbit and corresponding TD accretion rates $\Mdot$, in units of accreted mass per N-body time. For both panels, red solid line and blue dashed line represent $\rbh$ and $\Mdot$ respectively. Left panel corresponding to our fiducial model A05 with initial parabolic orbit, and the right panel is model C01 with initial orbital eccentricity $e = 0.3$ for comparison. Model C01 has been integrated to $t = 500$ because of the slower dynamical evolution of two SMBHs. Here we only pick up the last 200 time unit for easier comparison. According to Fig.~\ref{fig:TDrate}, the dynamical evolution of two SMBHs in moderately eccentric orbit is slower than parabolic orbit. Our integration result also shows that, for A05 the eccentricity at the end of integration keeps a constant level with $e\sim 0.7$, while in C01 it keeps around 0.05. Actually, almost all of our integrations with initial parabolic orbits result in high eccentricities for the SMBHBs at the end of integrations.

Fig.~\ref{fig:TDrate} indicates that there is a significantly enhanced TD accretion rate during two BHs getting close enough to form a bound binary. Thus we can empirically divide the entire evolution into three different phases. In phase I, $\rbh$, the separation of two BHs is far more larger than the influence radius of each BH, and $\Mdot$ keeps in a relatively constant low level. After two BHs getting close enough to have significant perturbation to each other, $\Mdot$ increases significantly and the system evolves into phase II. At the last phase III, a compact SMBHB has formed and $\Mdot$ drops down to a roughly constant lower level.

As shown in Fig.~\ref{fig:TDrate}, both model A05 and C01 have similar evolutions of TD accretion rates: there are roughly stable $\Mdot$ during phase I and III, and significantly enhanced $\Mdot$ in phase II. This result can be easily understood. In phase I, the evolution of $\Mdot$ is more or less the same as the case with single SMBH in isolate galaxy, because the perturbation from another nucleus is not significant. That corresponds to diffusive regime for the most of large galactic nuclei. However phase II is a stage with violent relaxation. There may be two factors contributing to the enhancement of TD accretion rate. First of all, stars bound to one SMBH suffer very strong perturbation from the companion SMBH, which will enhance the loss cone feeding. Secondly, due to triaxial stellar distribution and three body interaction in phase II, more stars can be scattered into the TD loss cone. As a result, the loss cone feeding during phase II is very efficient, which leads to significantly enhanced TD accretion rate.

\citet{liu13} have found that when the separation of two SMBHs is larger than $100\rinf$, the loss cone filling will be dominated by two-body relaxation and the TD accretion rate is identical to isolated single SMBH. While as their separation shrinks to $d \sim 100\rinf$, the TD accretion rate induced by perturbation from another galaxy will be comparable to the contribution from two-body relaxation. We find, however, for our models A05 and C01, that TD accretion rates are enhanced by perturbations only at around $d \lesssim 10\rinf$ or below. To understand this difference, we ran their code with the same parameters (galaxy mass, density profile and mass ratio of two galaxies) of our A05 model , but choosing realistic tidal radius and particle number. The result based on their model shows that the contribution from perturbation becomes higher than two-body relaxation around $d \lesssim 400\rinf$. But if we rescale in their model the TD accretion rate obtained from two-body relaxation to the artificially small particle number and large tidal radius as used in our N-body simulation (using the scaling relations found in Section~\ref{Extra_PI} and~\ref{Extra_PII}) the effect of perturbation becomes important at $d \lesssim 8\rinf$, consistent with our results. We will come back to this issue in Section~\ref{Extra}.

It is hard to distinguish which factor dominates the loss cone feeding in our simulation. Since we want to focus on the TDE rate evolution in this paper, the discussion about the contributions from triaxial stellar distribution and perturbation will be postponed to the next paper. In phase III, the SMBHB has been formed and evolved to a hard binary. Most of the stars bound to each BH have been ejected or disrupted by three body interaction. The TDE rate is dominated by stars outside the SMBHB orbit. Because of a triaxial stellar distribution around SMBHB, the TDE rate should be higher than a relaxed spherical distribution system like single galaxy. However, according to the results in left panel of Fig.~\ref{fig:TDrate}, the separation of two SMBHs is close to $\rt$ by the end of the integration, which will not happen so early in real systems with much smaller $\rt$. As a result, almost all of stars encountering the SMBHB in this stage will be disrupted. That will artificially suppress slingshot effects and enhance the TDE rate. For this reason, our results in late phase III tend to overestimate TD accretion rates.

\subsection{Dependence on particle number}
\label{Dep_N}

As mentioned in Section~\ref{mtd}, we can not afford a direct \nbody simulation with the same particle number as in a real galaxy. For a stellar system with embedded central SMBH in real galaxy, the total number of stars can be easily up to $10^9$, while direct \nbody simulations, even if using very powerful HPCs (high performance computers), can only manage $10^6 - 10^7$ particles right now. As a result, comparing to real galaxies, the relaxation time scale in our model should be shorter. Besides, if we set the initial mass of BH particle to $10^{-3}M$ and assume every star particle has equal mass $M/N$, the mass ratio of BH to star in our simulation will be much smaller than reality. For this reason, it is necessary to investigate the dependence of our simulation results on particle number. The data are given in Table~\ref{tab:AllRes}, with model index A01, A02, A05 and A11.

\begin{figure*}
\includegraphics[width=3.2in,angle=0.]{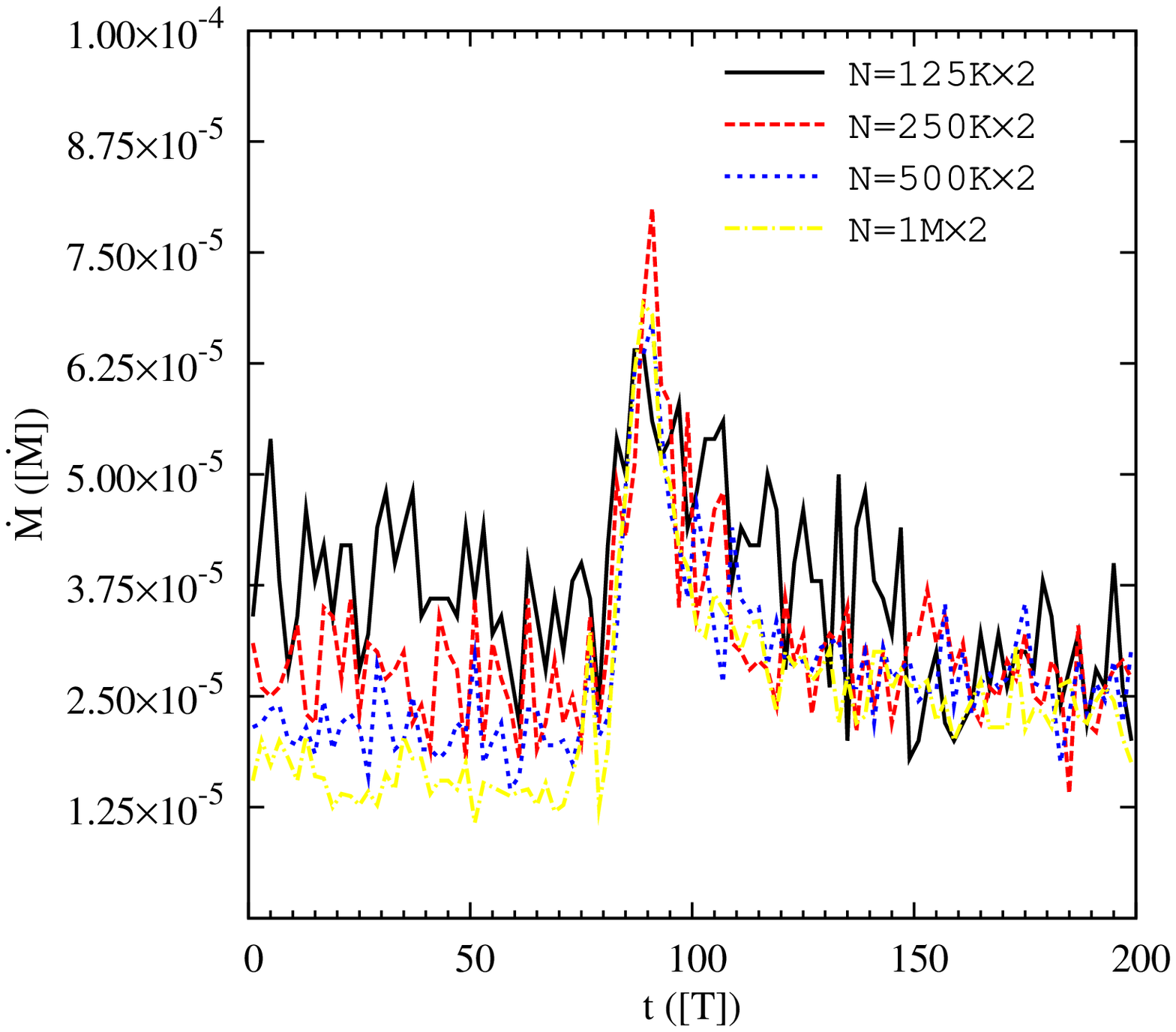}%
\hfill%
\includegraphics[width=3.2in,angle=0.]{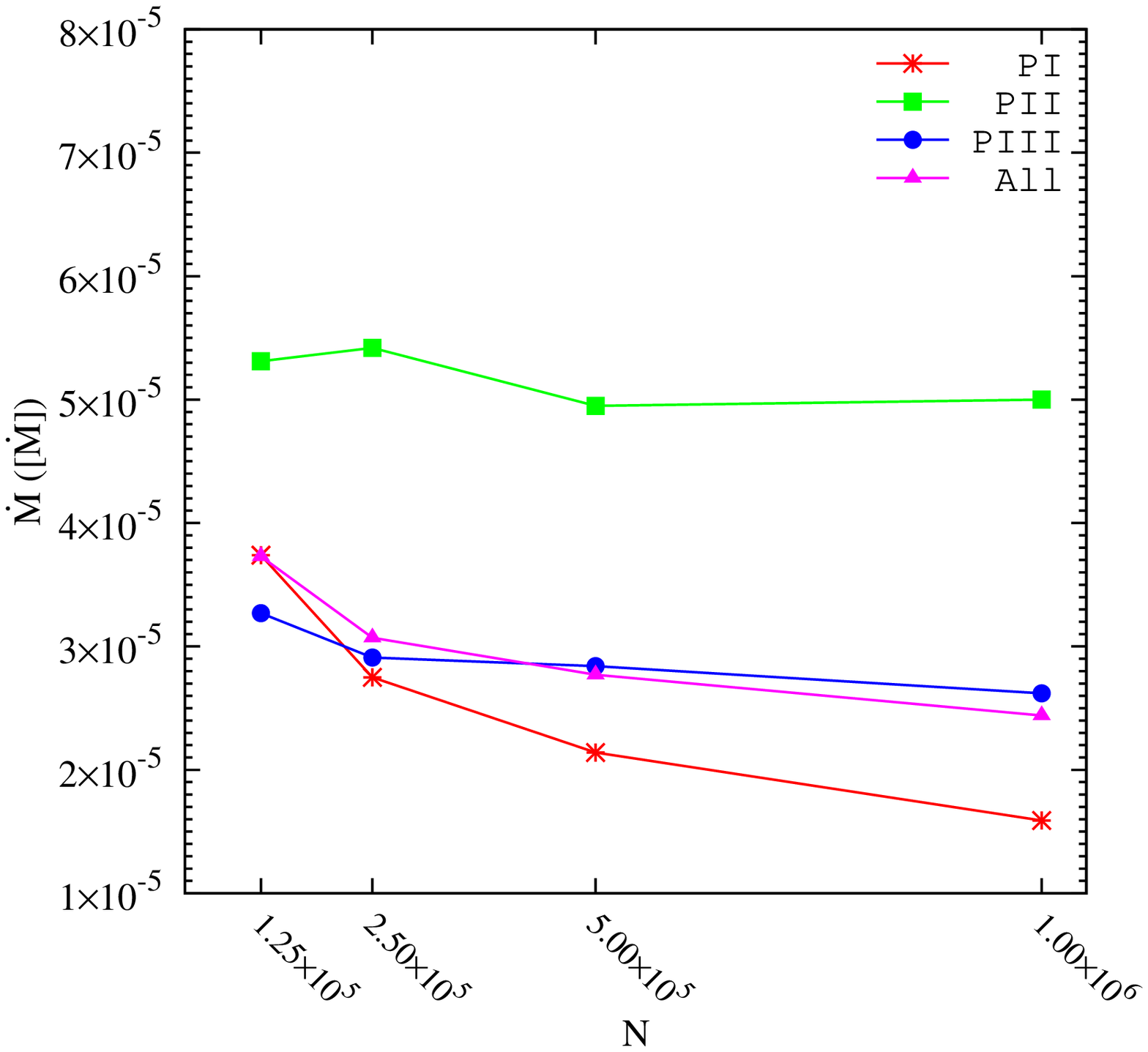}
\caption{Evolution of TD accretion rates and phase averaged TD accretion rates for different particle numbers $N$. Left panel: TD accretion rate evolution for different $N$. The \textit{y}-axis and \textit{x}-axis give TD accretion rate and evolved time in integration, respectively. Here TD accretion rate is in units of accreted mass per N-body time. Black solid, red dashed, blue dotted and yellow dash dotted lines represent different particle numbers, corresponding to different integration model A01, A02, A05 and A11, respectively. Right panel: Phase averaged TD accretion rates for different $N$. Here the red asterisks, green squares, blue circles and pink triangles correspond to averaged TD accretion rate during phase I, II, III and the entire integration, respectively. \label{fig:Ndep}}
\end{figure*}

Left panel of Fig.~\ref{fig:Ndep} shows $N$ dependence for the evolution of TD accretion rate $\Mdot$, in units of accreted mass per N-body time. Since the two SMBHs have roughly identical growth, all of our calculations for $\Mdot$ in this article are averaged to one SMBH. First of all, as represented by black solid line, the result of model A01 has relatively larger scatter comparing to others. We actually have some tests with even less particle numbers, and all of them can not give consistent results. As a result, we adopt $N = 125000$ per galaxy as the smallest particle number in our models. Similar to the results in Section~\ref{Evo_TD}, there are TD accretion rate peaks in phase II for all of integrations. In addition, for all of integrations during phase II, the $N$ dependence for the evolutions of TD accretion rate can be neglected. While in phase I and III there are significant and weak $N$ dependence, respectively. It is more intuitionistic in the right panel of Fig.~\ref{fig:Ndep}, which gives the $N$ dependence of averaged TD accretion rate in every phase. As represented in the right panel, the averaged TD accretion rate in phase I with red asterisks declines for increasing particle numbers, and the rate in phase III with blue circles also has a slight decrease. Phase II represented by green squares, conversely, roughly keeps a constant rate. That may because the loss cone feeding in phase I is dominated by two-body relaxation, which is N dependent. While the loss cone feeding during phase II is dominated by the perturbative effect of two SMBHs with their surrounding stars, the non stationary potential, and triaxial stellar distribution, all of which do not depend on $N$. Similar to phase II, both perturbation and triaxial gravitational potential also contribute to TD accretion rate during phase III, specially at the beginning of phase III. However, as the semi-major axis of SMBHB is decreasing, more and more stars close to SMBHB will be ejected or disrupted. Thus the contribution from the perturbation will be fading. At the same time, the contribution of the triaxial potential is also decreasing because the central region of the merger remnant is recovering to a spherical distribution. As a result, an ideal evolution of TD accretion rate in phase III will be dominated by two-body relaxation finally, which suggests a dependence on $N$ sooner or later. However, that may not happen in our simulations. Considering the large tidal radius we adopted here, the semi-major axis of SMBHB will be smaller than $\rt$ at some time, which may bring artificial effects. For this reason, our integrations terminate at $t = 200$, when the averaged separation of SMBHs close to $\rt$. In brief, the evolution of TD accretion rate in phase III is very complex, and we can only trace the evolution in its early stage.

\begin{deluxetable*}{crcrcrc}
    \tablewidth{0pt}
    \tabletypesize{\scriptsize}
    \tablecaption{Mass accretion rates for different models \label{tab:AllRes}}
    \tablehead{
    \colhead{No.} &
    \colhead{PI} &
    \colhead{$\dot{M}/Galaxy$} &
    \colhead{PII} &
    \colhead{$\dot{M}/Galaxy$} &
    \colhead{PIII} &
    \colhead{$\dot{M}/Galaxy$} \\
    \colhead{(1)} &
    \colhead{(2)} &
    \colhead{(3)} &
    \colhead{(4)} &
    \colhead{(5)} &
    \colhead{(6)} &
    \colhead{(7)}
    }
    \startdata

$A01$  & $0-83$ & $3.74\times10^{-5}$ & $83-109$ & $5.31\times10^{-5}$ & $109-200$ & $3.27\times10^{-5}$   \\
$A02$ & $0-84$ & $2.75\times10^{-5}$ & $84-102$ & $5.42\times10^{-5}$ & $102-200$ & $2.91\times10^{-5}$   \\
$A03$ & $0-87$ & $4.31\times10^{-6}$ & $87-101$ & $6.14\times10^{-6}$ & $101-200$ & $3.82\times10^{-6}$   \\
$A04$ & $0-86$ & $7.12\times10^{-6}$ & $86-107$ & $9.62\times10^{-6}$ & $107-200$ & $6.26\times10^{-6}$   \\
$A05$  & $0-83$ & $2.14\times10^{-5}$ & $83-104$ & $4.95\times10^{-5}$ & $104-200$ & $2.84\times10^{-5}$   \\
$A06$ & $0-80$ & $3.38\times10^{-5}$ & $80-102$ & $9.80\times10^{-5}$ & $102-200$ & $4.75\times10^{-5}$   \\
$A07$ & $0-108$ & $6.69\times10^{-6}$ & $108-135$ & $1.37\times10^{-5}$ & $135-200$ & $1.00\times10^{-5}$   \\
$A08$ & $0-65$ & $8.94\times10^{-5}$ & $65-82$ & $1.90\times10^{-4}$ & $82-200$ & $5.15\times10^{-5}$   \\
$A09$ & $0-92$ & $9.53\times10^{-6}$ & $92-143$ & $2.81\times10^{-5}$ & $143-200$ & $2.56\times10^{-5}$   \\
$A10$ & $0-73$ & $4.05\times10^{-5}$ & $73-90$ & $1.05\times10^{-4}$ & $90-200$ & $2.44\times10^{-5}$   \\
$A11$ & $0-82$ & $1.59\times10^{-5}$ & $82-102$ & $5.00\times10^{-5}$ & $102-200$ & $2.62\times10^{-5}$   \\

    \enddata

\tablecomments{ Col.(1): Model serial number in Table~\ref{tab:para}. Col.(2): Period of PI. Col.(3): TD accretion rate during PI. Col.(4): Period of PII. Col.(5): TD accretion rate during PII. Col.(6): Period of PIII. Col.(7): TD accretion rate during PIII.}
\end{deluxetable*}

\subsection{Dependence on tidal radius}
\label{Dep_rt}

\begin{figure*}
\includegraphics[width=3.2in,angle=0.]{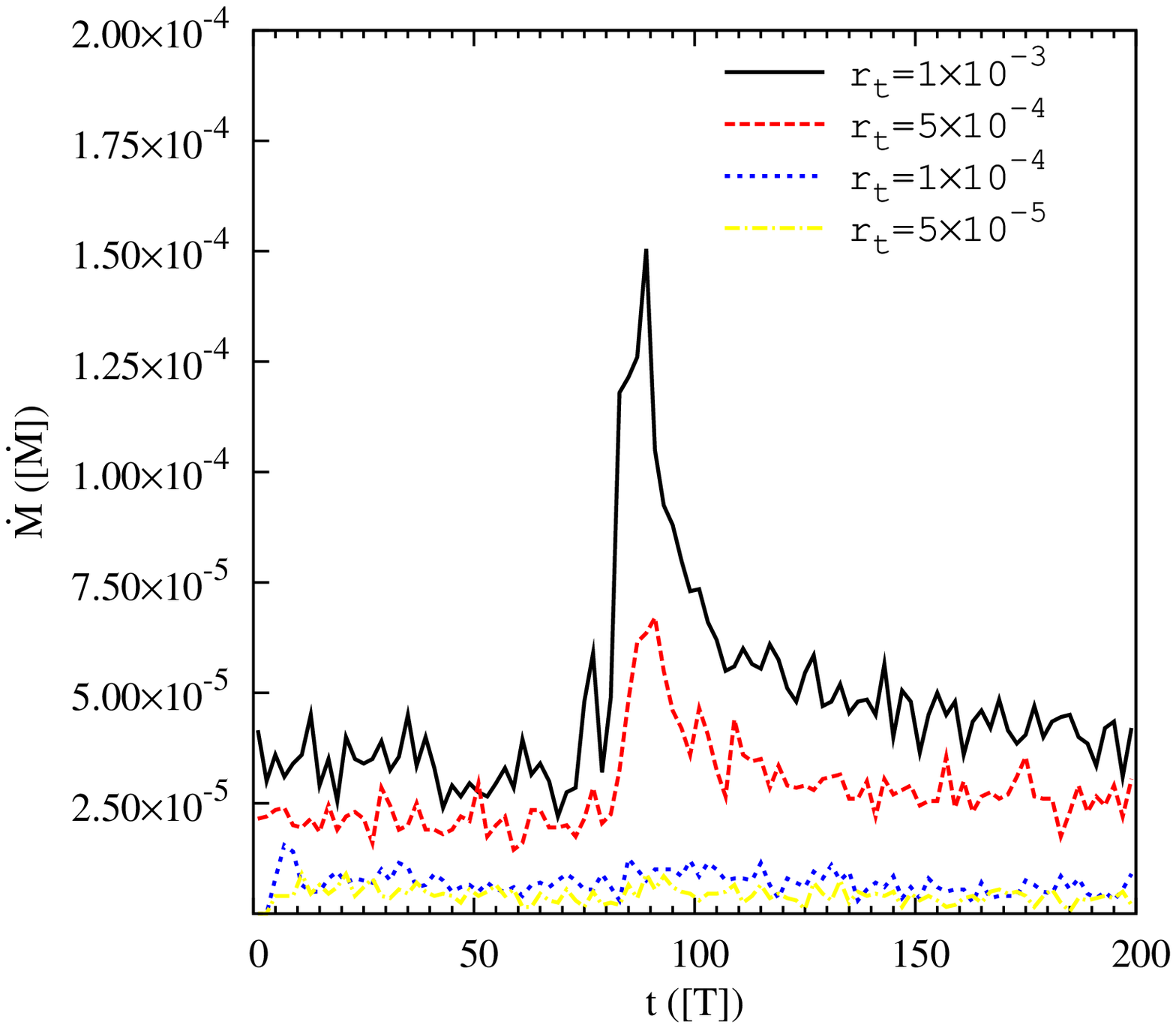}%
\hfill%
\includegraphics[width=3.2in,angle=0.]{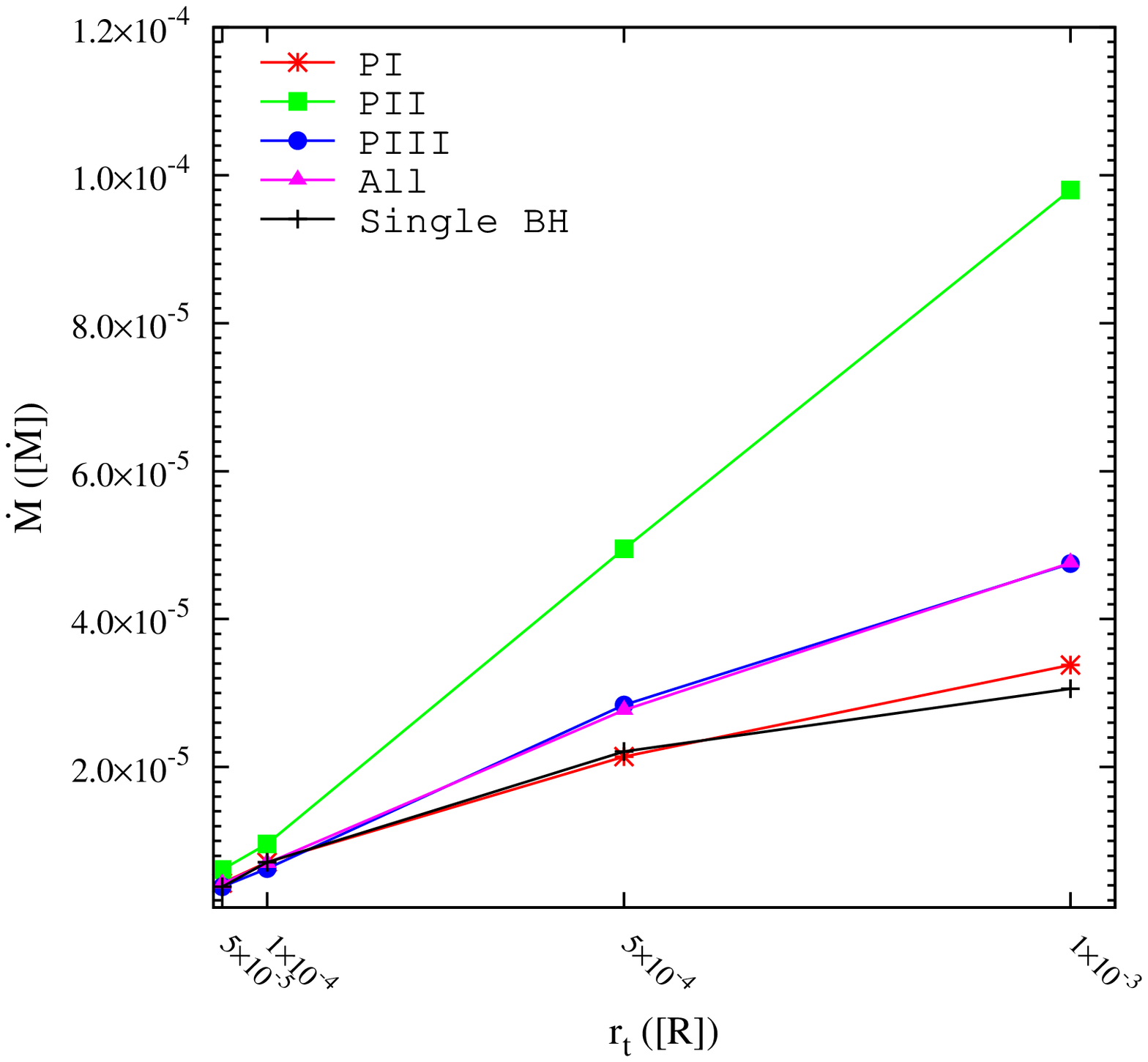}
\caption{Evolution of TD accretion rates for different $\rt$. Left panel: TD accretion rates as a function of time for different $\rt$. Black solid, red dashed, blue dotted and yellow dash dotted lines represent different integration model A06, A05, A04 and A03, respectively. The definitions of two axes are the same as in Fig.~\ref{fig:Ndep}. Right panel: Averaged TD accretion rates for different $\rt$ in different phases. Here different lines have the same meaning as in Fig.~\ref{fig:Ndep}. For comparison, the black plus line represents the results for single BH in isolated galaxy (model B01-B04. \label{fig:rtdep})}
\end{figure*}

In order to explore the dependence of our results on $\rt$, we have tried series integrations with different $\rt$. All the data are given in Table~\ref{tab:AllRes} with model index A03, A04, A05 and A06. Fig.~\ref{fig:rtdep} shows the influence of different $\rt$ to our results.  In principle, larger $\rt$ correspond to higher $\Mdot$. However, as shown in the left panel, when $\rt$ is smaller than $5\times10^{-4}$, there is no significant $\Mdot$ peak in phase II. According to our analytical estimation in Section~\ref{Extra} and simulation results shown in the right panel of Fig.~\ref{fig:rtdep}, there is a roughly linear dependence of $\Mdot$ on $\rt$. Meanwhile, this dependence in phase II is a power law with index less than one. For this reason, with $\rt$ decreasing, $\Mdot$ in phase II drops faster than phase I. For some small enough $\rt$, the contribution of phase II will be comparable to phase I, and the peak can be washed out. This is an artificial effect due to limited particle resolution in our simulations. In reality, due to much larger $N$, $\Mdot$ contributed by two-body relaxation in phase I should be smaller than we have in simulations. However, this effect is not crucial for our results. According to Table~\ref{tab:AllRes}, all of averaged TD accretion rates in phase II have notable enhancement even if $\rt=5\times10^{-5}$, which means the contribution from strong perturbation in phase II is still significant. A smaller $\rt$ may lead to equal accretion rate for phase I and II, but that does not mean our strategy is not suitable. For a proper scaling limit it is not sufficient just to keep one parameter fixed (e.g. $N$) and vary the other ($\rt$) to the realistically small value. In fact for every $N$ there is only a limited reasonable range of $\rt$. If $\rt$ is too small we will get too few events, while if it is too large we will get unphysically many events. The extrapolation can only be done simultaneously in $N$ and $\rt$, by carefully examining the scaling in both parameters, as we have done in Section~\ref{Extra}. In phase III, it seems that $\Mdot$ always keeps in a constant level which is slightly higher than phase I.

The right panel shows averaged TD accretion rates in different phases for different $\rt$. The correlation of TD accretion rate and $\rt$ in phase II can be well fitted by a linear correlation, implying that the loss cone is fed efficiently. In the reference \nbody simulations for isolated galaxy with single SMBH, the TD accretion rates represented by black plus line is similar to the results of phase I represented by red asterisk line. That is consistent with the expectation that TDE rate evolution in phase I is the same as a single BH in isolate spherical galaxy. Therefore, the TDE loss cone in phase I is empty and the loss cone feeding is dominated by two-body relaxation. Another interest thing is, the dependence of averaged TD accretion rate on $\rt$ in phase III, represented by blue circles, lies beside the results of phase I and II. It actually nearly overlap the pink triangles, the averaged result for all of three phases. The dependence of averaged TD accretion rate on $\rt$ in phase III is neither a linear relation like phase II, nor a relation close to single BH case, which indicates that the loss cone feeding in phase III is more efficient than phase I, but not as full as in phase II.

As shown in the left panel, the TD accretion rates for sufficiently large $\rt$ in phase I are slightly lower than that in phase III while significantly less than that in phase II. In other words, we have low TDE rates in phase I and slightly higher rates in phase III with relatively long period, and very high rates in phase II with very short period. In consideration of the short period of phase II, the contribution from phase II to the averaged TD accretion rates for all of three phases should not be dominant. For example, in model A05, the number of disrupted particles in phase II is only 1040, compared to a total of 5547 in the whole simulation. The contribution of phase I, II and III to the total TD rate is $\sim 32\%$, $\sim 19\%$ and $\sim 49\%$, respectively. According to the estimation in Section~\ref{obv}, the contribution of phase II is more than a quarter for our fiducial model. As a result, the rate in phase III is closer to the averaged result for all of three phases.

\subsection{Dependence on density profile and initial BH mass}
\label{Dep_gamma}

As discussed before, the numerical calculation capability limits the particle resolution, which can not provide a realistic high mass ratio of SMBHs to stars in direct \nbody simulation. Here we consider heavier SMBHs in the simulations to improve the situation. In addition to initial BH mass, it is also important to investigate the dependence of TDE rates on initial stellar density profile. Therefore we have carried out three integrations with $\Mbh = 0.0001, 0.001, 0.01$ and another three with $\gamma = 0.5, 1.0, 1.5$. Here we adopt the fiducial values for other model parameters. We do not have any tests  for $\gamma > 1.5$ because all of those integrations are prohibitively time consuming.

\begin{figure*}
\includegraphics[width=3.2in,angle=0.]{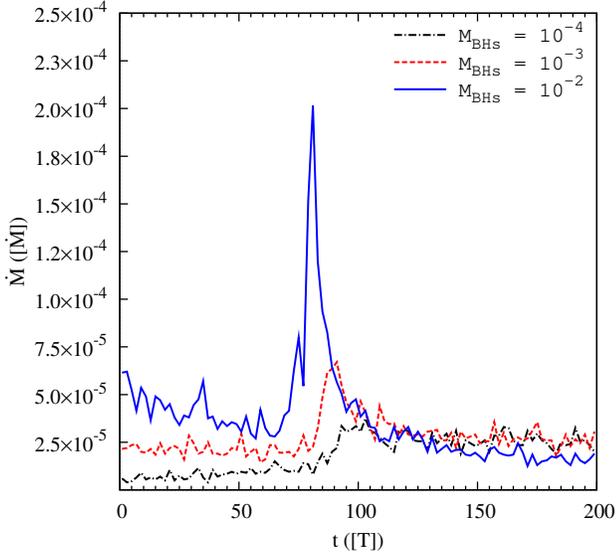}%
\hfill%
\includegraphics[width=3.2in,angle=0.]{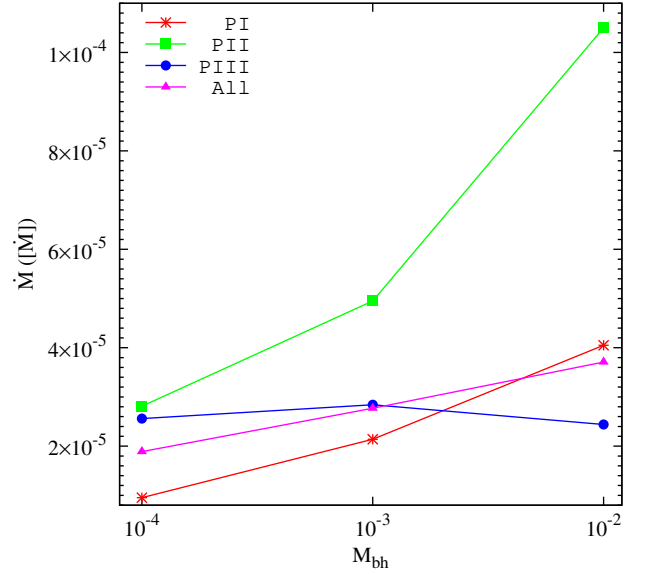}
\caption{Left panel: evolution of TD accretion rates for different initial SMBH mass. Black dash dotted, red dashed and blue solid lines are, respectively, for integration models A09, A05 and A10. Right panel: TD accretion rates as function of initial SMBH mass in different phases. Here different lines have the same meaning as in Fig.~\ref{fig:Ndep}.
\label{fig:Mdep}}
\end{figure*}

\begin{figure*}
\includegraphics[width=3.2in,angle=0.]{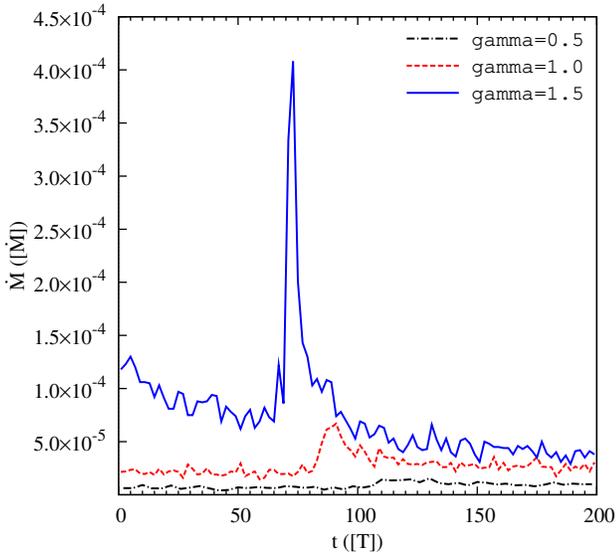}%
\hfill%
\includegraphics[width=3.2in,angle=0.]{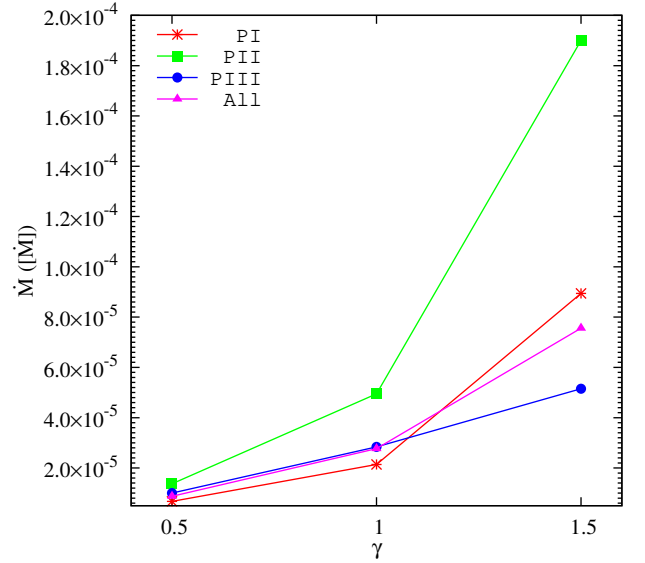}
\caption{Left panel: evolution of TD accretion rates for different initial stellar density index $\gamma$. Black dash dotted, red dashed and blue solid lines are for integration model A07, A05 and A08, respectively. Right panel: TD accretion rates as a function of $\gamma$ in different phases. Here different lines have the same meaning as in Fig.~\ref{fig:Ndep}.\label{fig:Gdep}}
\end{figure*}

The results are given in Figures~\ref{fig:Mdep} and \ref{fig:Gdep}. The left panels of Figures~\ref{fig:Mdep} and \ref{fig:Gdep} show that the enhancement of TD accretion rates and the starting time of phase II strongly depend on density distribution and initial BH mass. The larger $\Mbh$ and $\gamma$ are, the more significant enhancement of the TD accretion rates, and the earlier the staring time are. The duration of phase II also depends both on initial SMBH mass and density profile index $\gamma$. A system with more massive SMBHs or larger $\gamma$ usually corresponds to a shorter duration of phase II. More details can be found in Table~\ref{tab:AllRes}. It shows that model A07, A05 and A08 have different $\gamma$, and A09, A05 and A10 have different initial $\Mbh$. The influence of initial $\Mbh$ and $\gamma$ can be understood as following. In principle, more massive SMBH corresponds to larger influence radius $\rinf$ and plays stronger perturbation to its companion. Thus, SMBHs with larger initial $\Mbh$ would form bound system at larger separation and the formed SMBHB also evolves faster. While, for system with larger $\gamma$, the stellar distribution is more concentrated toward the center and the SMBH, and its TDE rates are enhanced much more significantly, leading to the faster growth of SMBH masses. In particular, at the transient time between Phase I and II, the differences of the BH mass between models A05 ($\gamma=1.0$) and A07 ($\gamma=0.5$), and between A08 ($\gamma=1.5$) and A05 are, respectively, $\sim1.6$ and $\sim2.5$ times. Heavier SMBH mass plays stronger perturbation to the stars around the companion SMBH, leading to faster ejection and more frequent TDEs and thus shorter duration of phase II. But this effect may not be significant in reality because the mass growth of SMBHs in our \nbody simulation is artificially enlarged. It is interesting to notice that for the heaviest initial SMBH mass $\mbh = 0.01$ and the largest density profile index $\gamma = 1.5$, the TD accretion rates evolution in Phase I have slight decrease until the start of Phase II. It may be due to a problem of the set-up of initial density distribution. Because the systems for model A10 and A08 have much longer relaxation timescale in the bound regions and at the influence radius of SMBH, our limited computer resources prohibit us to get the systems well relaxed before the integrations start.

The right panels of Figures~\ref{fig:Mdep} and \ref{fig:Gdep} give the averaged TD accretion rates at different phases as a function of initial $\Mbh$ and $\gamma$, respectively. The averaged TD accretion rates in phase I and II significantly correlate with the initial BH mass and stellar distribution of galactic nucleus.  In phase II, the TD accretion rate for model of $\gamma=1.5$ (Model A08) is about fourteen times higher than that for $\gamma =0.5$ (Model A07).  It is expected that the TD accretions for $\gamma > 1.5$ is even higher. In addition, the rate in phase II also strongly depends on initial mass of SMBH, and the rate for $\Mbh = 0.01$ (Model A10) is about four times that for $M_{\rm BH} = 0.0001$ (Model A09). In phase III, the averaged TD accretion rate weakly depends on $\gamma$ and nearly independent of $\mbh$.

\section{Extrapolation to Real Galaxies}
\label{Extra}

To apply our simulation results in real systems of galaxy mergers, we have to do extrapolations.  To obtain TDE rates, one needs to integrate the Equation~(\ref{eq:TDNRT}). Since TDE rate is dominated by the contribution of stars around the critical radius $\rcrit$,  we can estimate TDE rate
\beq
\Gamma \sim \left. \frac{r^3\theta^{2}_{\rm{D}}\rho(r)}{t_{\rm d}\mstar}\right|_{r=r_{\rm{crit}}}
\label{eq:TDrate}
\eeq
 \citep{fra76,bau04b}, where for simplicity all the stars are assumed to have equal mass ($\mstar=\msun$). Now the key point is
to estimate $\rcrit$. As we will show in the following sections, $\rcrit$ changes with the dynamical evolution of SMBHs during
galaxy mergers. For order of magnitude, we estimate the averaged TDE rates with typical critical radius $r_{\rm crit}$
for each phase.

\subsection{Extrapolation for phase I}
\label{Extra_PI}

In phase I, the separation of two SMBHs is much larger than the influence radius of SMBH (typically $r_{\rm BH} \gg \rinf$).
As was shown by \citet{liu13}, the perturbation and torque of the companion galactic core is small in this stage. The typical TDE rate
can be estimated as to that of an isolated galaxy with a single central SMBH. For a spherical isotropic system, two-body
relaxation dominates the diffusion process and $\thetaD^2 \sim \thetatwo^2$ in Equation~(\ref{eq:theta_d2}) with
\beq
\thetatwo^2 \backsimeq \frac{t_{\rm d}}{t_{\rm r}}
\label{eq:thetatwo_2}
\eeq
\citep{fra76}, where the dynamical time of star is $t_{\rm d}\backsim \sqrt{r^3/(G\Mbh)}$ within the influence radius and the two-body
relaxation time scale $t_{\rm r}$ is
\beq
t_{\rm r} \thickapprox \frac{0.34\sigma^3}{G^2\rho m_* \ln \Lambda}
\label{eq:t_r}
\eeq
\citep{binn08}. Here $\ln \Lambda$ is the Coulomb logarithm and $\sigma$ is one dimensional velocity dispersion of stars. For $r\la \rinf$,
we have $\sigma \thickapprox \sqrt{G\Mbh/r}$. The Coulomb logarithm is approximately
\beq
\ln \Lambda \thickapprox \ln \left(\frac{\Mbh}{2m_*}\right)
\label{eq:CL}
\eeq
\citep{pret04}.

From Equation~(\ref{eq:theta_lc2a}), (\ref{eq:thetatwo_2}), and (\ref{eq:t_r}) and the assumption $\rho(r) = \rho_0 (r/r_0)^{-\gamma}$, we have the typical critical radius of Phase I

\begin{eqnarray}
r_{\rm{crit}}&\approx &\left(0.23\Mbh^2\frac{\rt}{m_*\rho_0 r_0^\gamma\ln\Lambda}\right)^{\frac{1}{4-\gamma}} \nonumber \\
&\propto& \Mbh^{\frac{2}{4-\gamma}}\left(\frac{N}{\ln\Lambda}\right)^{\frac{1}{4-\gamma}}\rt^{\frac{1}{4-\gamma}}.
\label{eq:rcrit}
\end{eqnarray}
Thus, Equation~(\ref{eq:TDrate}) gives the mass accretion rate due to TD
\beq
\dot{M} = \Gamma m_* \sim \frac{r^3\theta^{2}_{\rm{D}}\rho(r)}{t_{\rm d}}\mid_{r=r_{\rm{crit}}}
\label{eq:Mdot}
\eeq
and

\begin{eqnarray}
\dot{M}&\approx & 2.94\times 0.23^{\frac{9-4\gamma}{8-2\gamma}}G^{\frac{1}{2}}\left(\rho_0r_0^\gamma\right)^{\frac{7}{8-2\gamma}}\Mbh^{\frac{6-5\gamma}{8-2\gamma}} \nonumber \\
&\times&\left(\frac{N}{\ln\Lambda}\right)^{-\frac{2\gamma-1}{8-2\gamma}}\rt^{\frac{9-4\gamma}{8-2\gamma}} \nonumber \\
&\propto& \Mbh^{\frac{6-5\gamma}{8-2\gamma}}\left(\frac{N}{\ln\Lambda}\right)^{-\frac{2\gamma-1}{8-2\gamma}}\rt^{\frac{9-4\gamma}{8-2\gamma}}.
\label{eq:TDRs}
\end{eqnarray}

We assumed that $\rho_0 r_0^\gamma$ is independent of $\Mbh$, $N$ and $\rt$. This is different from the assumption had been used by \citet{fra76}, which assumed that $\rho_0 = \rho_{\rm{h}}$ and $r_0 = \rinf$, with $\rho_{\rm{h}}$ representing the density at influence radius. $\rinf$ depends on $\Mbh$ by its definition $\rinf \simeq G\Mbh/\sigma_*^2$. Here $\sigma_*^2$ denotes one dimensional stellar velocity dispersion, which they considered to be constant outside $\rinf$. As a result, they argued that TD accretion rate should sensitively depend on BH mass. But $\sigma_*$ is a strong function of radius in cuspy galaxy models such as Dehnen model we adopted here \citep{trem94, merr07}. Therefore, we simply consider $\rho_0 r_0^\gamma$ as a undetermined coefficient, and estimate the TD accretion rate in reality by fitting and extrapolating our simulation results. Our results based on direct \nbody simulation should be more reliable. For this reason, we prefer to choose Equation~(\ref{eq:TDRs}) as our extrapolation solution.

As mentioned in Section~\ref{mtd}, our model has artificially enhanced BH accretion, which may be not accurate enough to estimate TD accretion rate in phase I. Considering the analyze above and our simulation results, we believe this effect is not significant. Our results show that, by the end of phase I, $\Mbh$ and $\rinf$ has been increased up to $\sim 2.6$ and $\sim 1.6$ times, respectively. For most of our models, at beginning, the relaxation time scale around $r=\rinf$ for each galaxy is longer than the duration of phase I. For simplicity, we can assume that this relation could persist during the entire phase I. That means the mass growth of SMBHs will only impact very central region within phase I, which can not efficiently change TD accretion rate. Combining with a non-growing tidal radius adopted in the simulation, the artificial mass growth of SMBH do not have enough time to make serious effects during phase I. In order to check this speculation, we have calculated the stellar dispersion and density evolution of model B03, and also two additional similar integrations with $N=250K$ and $N=125K$. The comparisons between $t=0$ and $t=80$ show that there is no significant change for these two parameters around $\rinf$. Our analysis on $\dot{M}$ - $\Mbh$ dependance also indicates that, along with the growth of SMBH, $\dot{M}$ roughly keeps constant in phase I. For this reason, we believe that the mass growth of SMBH can not significantly change the TDE rate in phase I. However, it does not means the influence of initial $\Mbh$ is negligible.  Because the initial stellar distribution is sensitively depends on initial $\Mbh$.

For fixed $\Mbh$ and $N$, we have relation
\beq
\dot{M} \propto \rt^{\frac{9-4\gamma}{8-2\gamma}} .
\label{eq:TDRrt}
\eeq

\begin{figure*}
\includegraphics[width=2.4in,angle=270.]{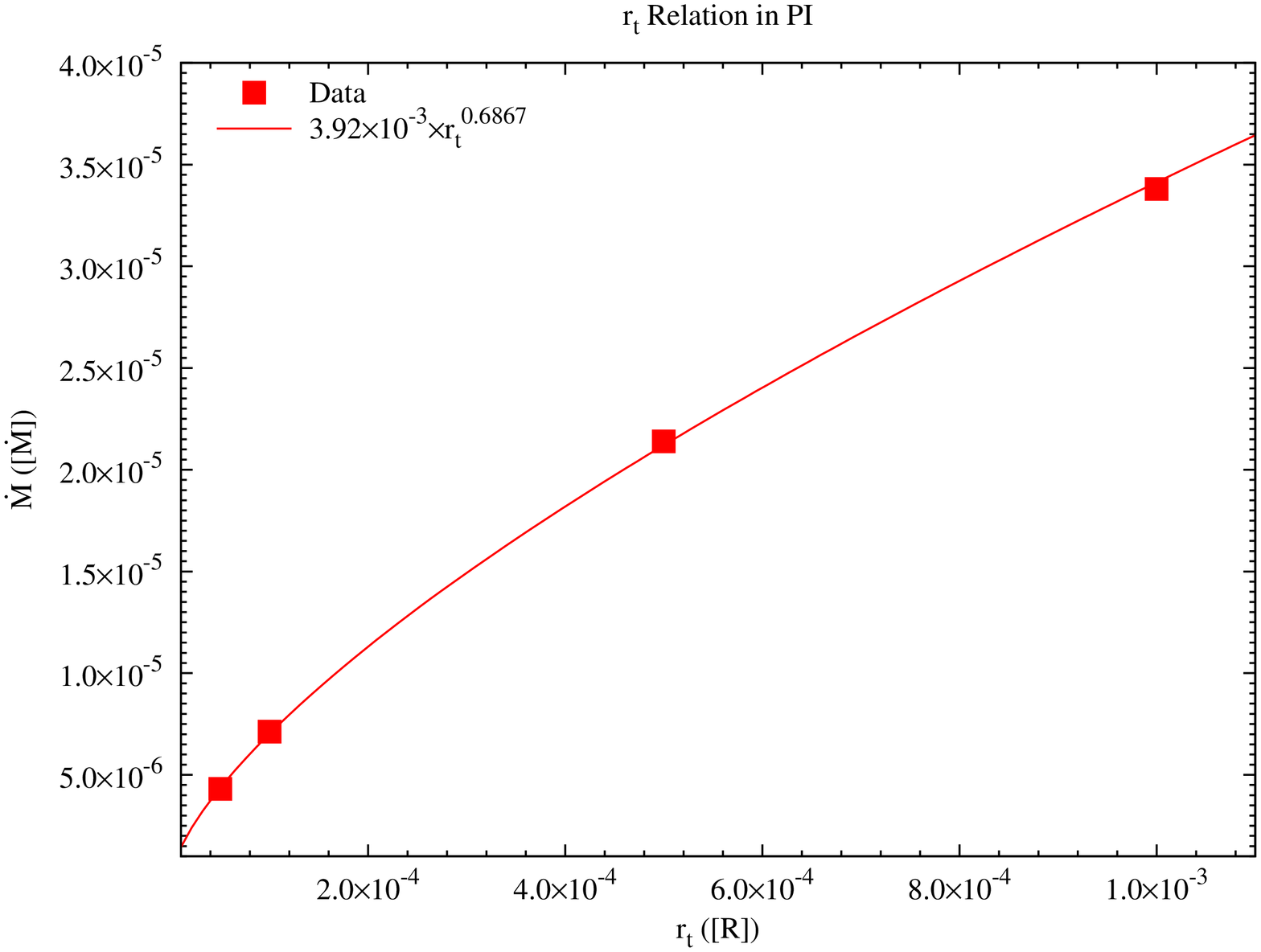}%
\hfill%
\includegraphics[width=2.4in,angle=270.]{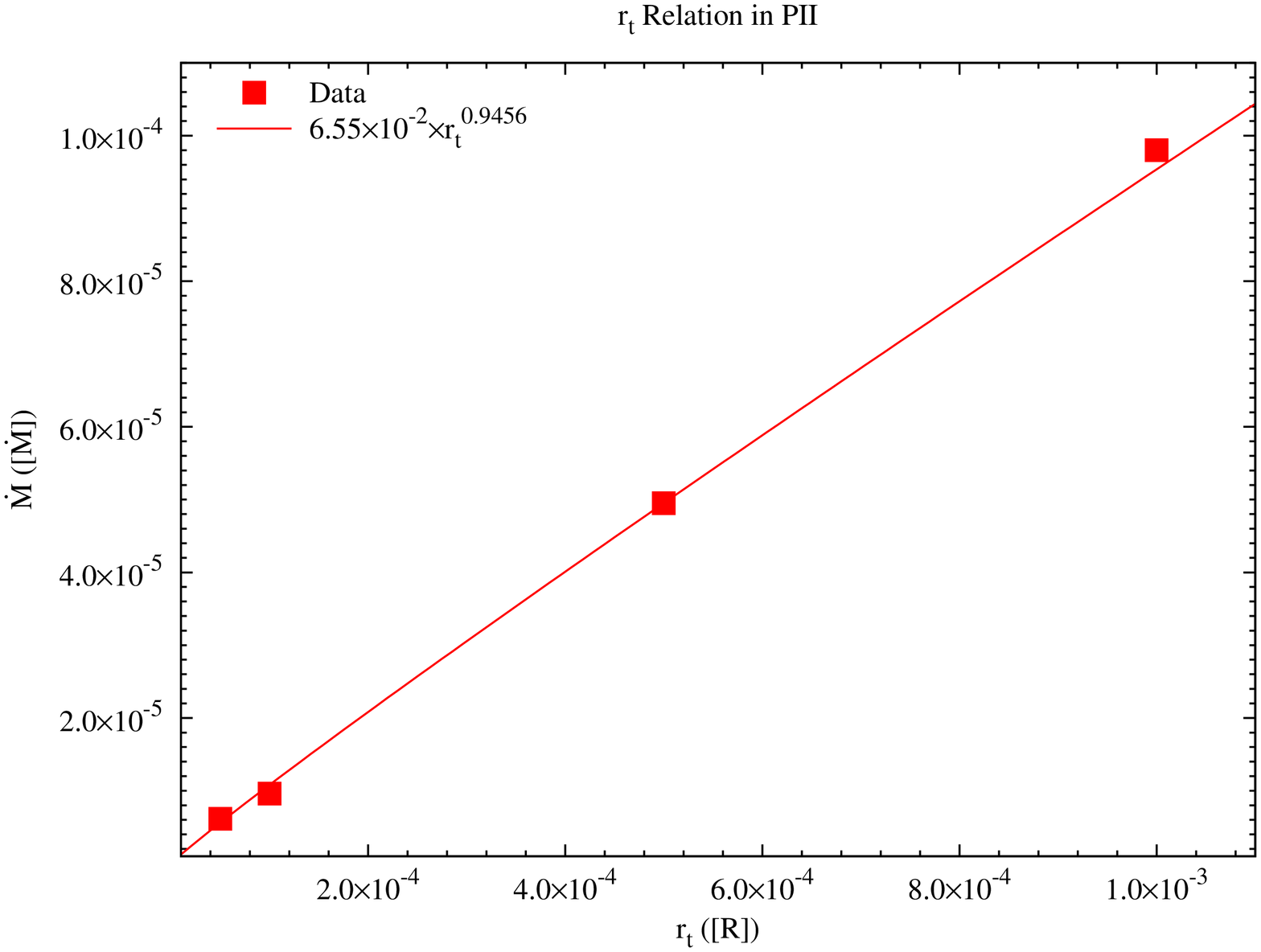}%
\hfill%
\includegraphics[width=2.4in,angle=270.]{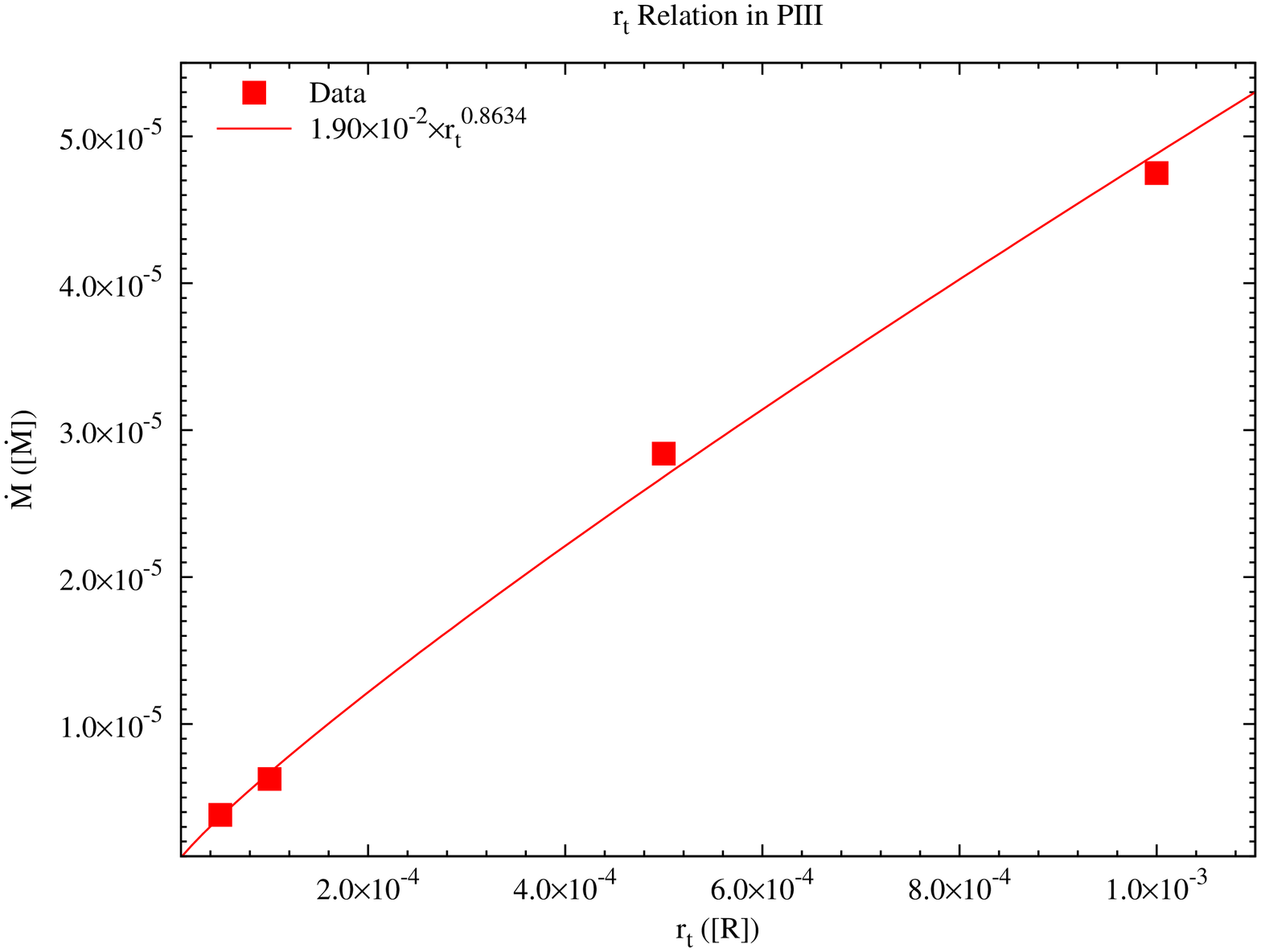}%
\hfill%
\includegraphics[width=2.4in,angle=270.]{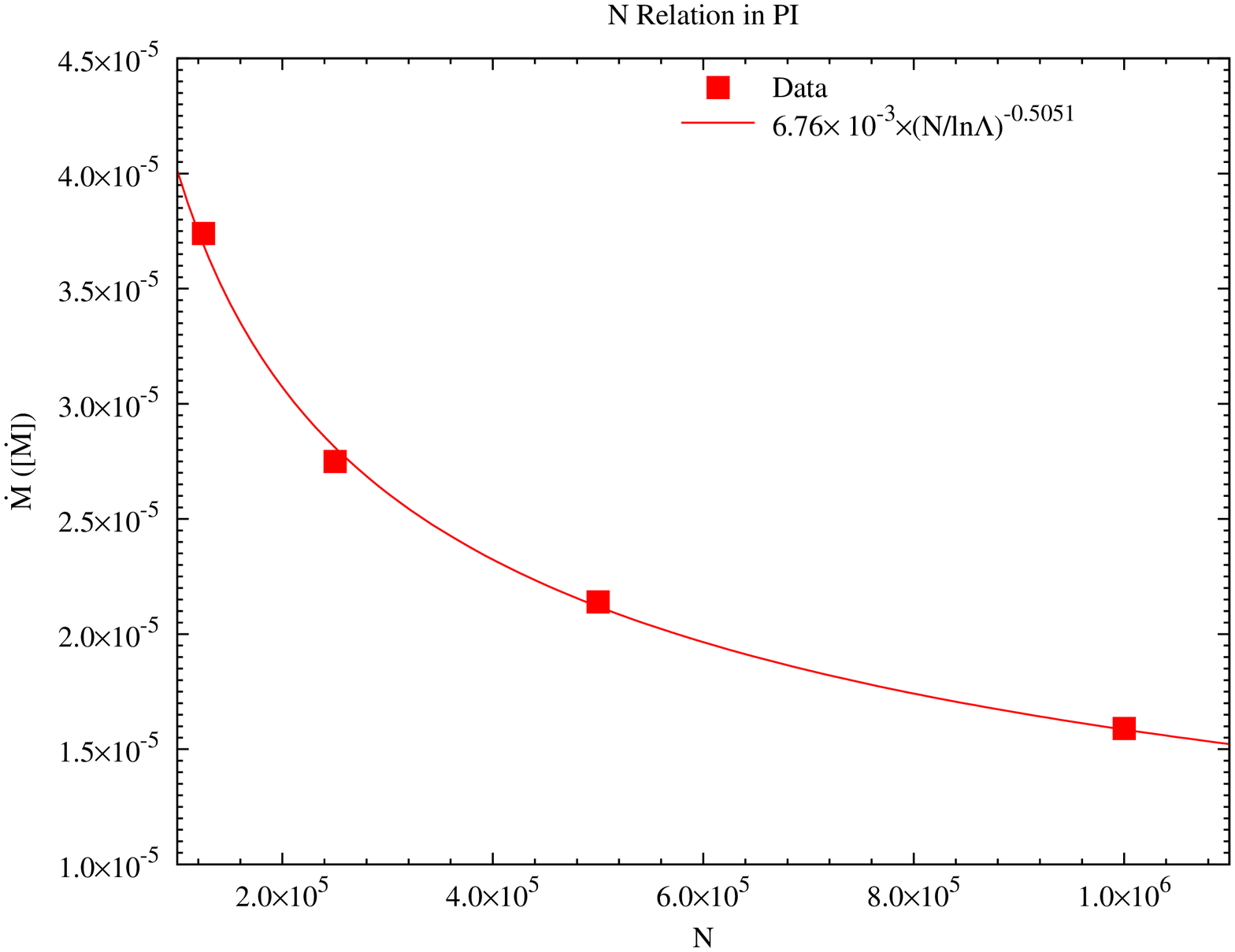}
\caption{Fitting results for $\rt$ and N dependence. Here the red solid squares represent the integration results in our simulations, and the red solid lines represent the fitting results according to formula~(\ref{eq:TDRs}), (\ref{eq:TDRrt}) and (\ref{eq:TDRsPII}), or a power law relation. The upper left, upper right and bottom left panel show the $\rt$ fitting result for phase I, II and III, individually. The bottom right panel gives the N fitting result for phase I.
\label{fig:fit}}.
\end{figure*}

Although Equations~(\ref{eq:TDRs}) and (\ref{eq:TDRrt}) are valid for $r_{\rm{crit}}<\rinf$, it can be expected that the power-law relationship between mass accretion rate and $\rt$ or $N$ should be a good approximation even for $r_{\rm{crit}} \sim \rinf$. Fig.~\ref{fig:fit} shows the simulated TD accretion rates of Phase I as functions of $r_{\rm t}$ (the upper left panel) and particle number $N$ (bottom right panel), and the fits with power-laws as suggested by Equations~(\ref{eq:TDRs}) and (\ref{eq:TDRrt}). The TD accretion rates, which obtained by direct \nbody simulations, as functions of $r_{\rm t}$ and $N$ are well fitted by power-laws $\dot{M} \simeq 3.92\times 10^{-3} r_{\rm t}^{0.6867}$ and $\dot{M} \simeq 6.76\times 10^{-3} (N/\ln\Lambda)^{-0.5051}$, respectively. The accretion rate is in our simulation units $[\Mdot]$. With these fitting results, as well as Equations~(\ref{eq:TDRs}), we can derive $\gamma\sim 1.33$ and $\gamma\sim 1.67$, respectively. We consider these two results as consistent, given the approximations used here.

Actually, according to Equation~(\ref{eq:TDRs}), with fixed $\Mbh$, we can derive
\beq
\ln\dot{M}\approx \ln A - \frac{2\gamma-1}{8-2\gamma}\ln\left(\frac{N}{\ln\Lambda}\right) + \frac{9-4\gamma}{8-2\gamma}\ln\rt,
\label{eq:lnTDRs}
\eeq
where
\beq
A = 2.94\times 0.23^{\frac{9-4\gamma}{8-2\gamma}}G^{\frac{1}{2}}\left(\rho_0r_0^\gamma\right)^{\frac{7}{8-2\gamma}}\Mbh^{\frac{6-5\gamma}{8-2\gamma}}.
\eeq
Based on Equation~(\ref{eq:lnTDRs}) we can make a two-dimensional fitting, which gives results consistent with those we got above
\beq
\dot{M} \sim 1.26\times \left(\frac{N}{\ln\Lambda}\right)^{-0.506}\rt^{0.686}.
\eeq
With Equation~(\ref{eq:scalingMdot}), we can finally derive the TD accretion rate for real galaxy

\begin{eqnarray}
\dot{M} &\sim & 0.387\times\left(\frac{M}{10^{11}\msun}\right)^{3/2}\left(\frac{r_{1/2}}{1\kpc}\right)^{-2.186} \nonumber \\
        &\times& \left(\frac{N}{\ln\Lambda}\right)^{-0.506}\left(\frac{\rt}{10^{-6}\pc}\right)^{0.686} \msun/\yr .
\end{eqnarray}

Now we can easily extrapolate our simulation results of phase I to real galaxies. For our fiducial galaxy model, the averaged TD accretion rate in phase I should be $\sim 7\times10^{-6} \accrate$. For comparison, we also have calculated the averaged TDE rates for galaxies similar to M32 or M105. For a M32 like galaxy, according to observations, we adopt $r_{1/2}\sim 0.1\kpc$ and $\Mbh\sim 3\times10^6\msun$\citep{wan04,capp06}. Here we assume that the half mass radius of the galaxy is roughly equal to effective radius. In order to keep consistent with our simulation model, we set $M\sim 3\times10^9\msun$ and $\gamma = 1.0$. As a result, our extrapolation shows that the averaged TD accretion rate in phase I is $\sim 4\times10^{-5}\accrate$. For more massive galaxy similar to M105, with $r_{1/2}\sim 2\kpc$, $\Mbh\sim 10^8\msun$, $M\sim 10^{11}\msun$ and $\gamma = 1.0$, the averaged TDE rates in phase I is $5\times10^{-6}\accrate$.

Since the averaged TDE rates in phase I of merger should be similar to a normal galaxy with single SMBH, it will be interesting to compare our results with the estimations of normal galaxies. For a merging system compose two galaxies similar to M32, the averaged TDE rate in phase I is an order of magnitude lower than that argued by \citet{wan04}. That may because \citet{wan04} used a density slope index of $\gamma = 2$, which is larger than we have for $\gamma=1$. As Fig.~\ref{fig:Gdep} and Equations~(\ref{eq:TDRs}) and (\ref{eq:TDRrt}) show, TDE rates depends on $\gamma$. By fitting the dependence of averaged TD accretion rates on $\gamma$, which is given in the right panel of Fig.~\ref{fig:Gdep}, we get a power-law index $\sim 2.29$. After simply scaling to $\gamma=2$, we have $\dot{M} \simeq 2\times10^{-4}\accrate$ for M32. Considering that there are differences for the results obtained through different methods, this result is consistent with their estimation \citep{alex12}.

\subsection{Extrapolation for phase II}
\label{Extra_PII}

In phase II, the separation of two SMBHs evolves from $\sim20\rinf$ to $\sim3\times10^{-2}\rinf$, where $\rinf \sim 0.1$ for an initial stellar distribution with $\gamma=1.0$. In a real galaxy merger, that corresponds to the evolution of two SMBHs from formation of bound SMBHB to hard binaries. Because of the strong perturbation by the companion galactic core and of the triaxiality in stellar distribution, we have $\theta_{\rm p}^2 \gg \theta_{\rm 2}^2$, $\theta_{\rm p}^2 \gg \theta_{\rm c}^2$  and $\theta_{\rm d}^2 \sim \theta_{\rm p}^2$ in Equation~(\ref{eq:theta_d2}).

As was shown by \citet{liu13}, the companion SMBH together with the bound star cluster would exert a strong tidal torque to the stars around the main SMBH with distance $r$. If $M_{\rm p}$ is the total mass of the perturber and $\rbh$ is the distance between the two SMBHs, the tidal torque is approximately
\beq
T_{\rm p} \sim \frac{{\rm G} M_{\rm p}r^2}{\rbh^3}
\label{eq:Tp}
\eeq
for $r\ll \rbh$. Thus the changes of the angular momentum of a star can be up to a time scale $t_{\rm \omega} = \min(t_{\rm d,M_p}, t_{\rm prc})$, with $t_{\rm d,M_p}$ and $t_{\rm prc}$ are, respectively, the dynamical timescale of the perturber and the stellar apsidal precession timescale. For $t>t_{\rm \omega}$, we can estimate the averaged change of the angular momentum of star on dynamical timescale $t_{\rm d}(r)$ \citep{liu13}.

\beq
J_{\rm p}^2 = T_{\rm p}^2t_{\rm \omega}t_{\rm d}(r)
\label{eq:J_p2}
\eeq

Because $\theta_{\rm p}^2=J_{\rm p}^2/J_{\rm c}^2$, we have
\beq
\cases{
\theta_{\rm p}^2 = \frac{J_{\rm p}^2}{(r\sigma)^2} , \quad {\rm for~r\gtrsim \rinf} \cr
\cr
\theta_{\rm p}^2 = \frac{J_{\rm p}^2}{{\rm G}\Mbh r} ,  \quad {\rm  r< \rinf}.
}
\label{eq:theta_p2}
\eeq
Where we have assumed a constant stellar velocity dispersion for $r\gtrsim \rinf$ and $\sigma^2 = G\Mbh/\rinf$.
From Equations~(\ref{eq:theta_lc2a}) and (\ref{eq:theta_p2}) and with the assumption of $r \la \rinf$, we obtain
the critical radius
\beq
\rcrit \simeq 0.93{\rm G}^{-1/11}\mbh^{3/11}M_{\rm p}^{-4/11}\rbh^{12/11}t_{\rm \omega}^{-2/11}\rt^{2/11}.
\label{eq:rcritPII}
\eeq
Table \ref{tab:AllRes} and discussions in Section~\ref{Dep_N} imply that the typical value of the separation $\rbh$
 in Phase II depends neither on $N$ nor $t_{\rm \omega}$. Because $M_{\rm p}$ depends on $\rbh$, we have $\rcrit\propto
 \rt^{2/11}$, which is independent of $N$. From Equations~(\ref{eq:theta_lc2a}), (\ref{eq:Mdot}) and (\ref{eq:rcritPII}), we have

\begin{eqnarray}
\dot{M}&\approx & \frac{2}{3}\rho_0 r_0^\gamma ({\rm G}\mbh)^{1/2}\rcrit^{1/2-\gamma}\rt \nonumber \\
&\approx & \frac{2}{3}\times 0.93^{\frac{1-2\gamma}{2}}\rho_0r_0^\gamma G^{\frac{5+\gamma}{11}}M_p^{\frac{4-2\gamma}{11}}\rbh^{\frac{6-12\gamma}{11}}t_{\rm \omega}^{\frac{2\gamma-1}{11}}\nonumber \\
&\times&\mbh^{\frac{7-3\gamma}{11}}\rt^{\frac{12-2\gamma}{11}} \nonumber \\
&\propto & \mbh^{\frac{7-3\gamma}{11}}\rt^{\frac{12-2\gamma}{11}}.
\label{eq:TDRsPII}
\end{eqnarray}

Equation~(\ref{eq:TDRsPII}) implies that the TD accretion rate in Phase II is approximately a power law functions of $M_{\rm BH}$ and $r_{\rm t}$. The upper right panel of Fig.~\ref{fig:fit} shows that our \nbody simulation results for $\dot{M}$--$r_{\rm t}$ in Phase II can be well fitted with a power law function with $\dot{M} = 6.55 \times 10^{-2} \, r_{\rm t}^{0.9456}$, which is consistent very well with Equation~(\ref{eq:TDRsPII}) for $\gamma=1.0$. We did not subtracted the contribution of two-body relaxation in phase II. According to Fig.~\ref{fig:Ndep} and our analysis above, the TD accretion rate in phase II does not depend sensitive to $N$, therefore we conclude that the contribution of two-body relaxation here is negligible. Similar to Section~\ref{Extra_PI}, we can estimate the TD accretion rate for a real galaxy in phase II:
\begin{eqnarray}
\dot{M} &\sim & 1.17\times 10^{-4} \left(\frac{M}{10^{11}\msun}\right)^{3/2} \nonumber \\
&\times&\left(\frac{r_{1/2}}{1\kpc}\right)^{-2.4456}\left(\frac{\rt}{10^{-6}\pc}\right)^{0.9456} \msun/\yr.
\end{eqnarray}

With this fitted power-law relation, we can extrapolate our simulation results in phase II to a real system of galaxy merger. For our fiducial model of SMBH mass $M_{\rm BH} = 4\times 10^7 M_\odot$, the averaged TD accretion rate in phase II is $\sim 2\times10^{-4} \accrate$ with a peak TD accretion rate $\sim 5.5\times10^{-4} \accrate$. The TD accretion rate for Phase II is about 30 times for average and about 80 times at peak higher than the rate in Phase I.  For a M32 like galaxy with $M_{\rm BH} = 3\times 10^6 M_\odot$, the averaged TD accretion rate for Phase II is $5\times10^{-4}\accrate$, and the peak rate is $\sim 1.4\times10^{-3} \accrate$, which is about 13 times for average or 35 times at the peak higher than the rate in Phase I. For more massive galaxy similar to M105 with $M_{\rm BH} \sim 10^8 M_\odot$, the TD accretion rate is $2\times10^{-4}\accrate$ with peak $\sim 5.4\times10^{-4} \accrate$.

The TDE rates given here is averaged over a range of the SMBH separation from about $20 \rinf$ to $3\times 10^{-2} \rinf$. \citet{liu13} analytically investigated the TDE rates as a function of SMBH separation in galaxy mergers for $r_{\rm BH} \sim 10^3 r_{\rm b}$ -- $2r_{\rm b}$ with $r_{\rm b} \sim \rinf$, which overlap with the early stage of Phase II. Their Fig.~4 shows that the TDE rates can be enhanced by up to $\sim 200$ at $r_{\rm BH} \sim 2\rinf$ over the rates at $r_{\rm BH} \gg 100 \rinf$ for $\gamma=1.75$, $M_{\rm BH} =10^7 M_\odot$ and SMBH mass ratio $q=1$. That are, respectively, about 7 times and 3 times larger than the averaged and peak values for our fiducial model. The difference may be because they used a deeper density profile with $\gamma = 1.75$. As it was shown in Section~\ref{Dep_gamma}, a larger $\gamma$ usually corresponds to a larger TDE rate. Similar to the $\gamma$ extrapolation for phase I, if we fit the averaged TD accretion rates for different $\gamma$ in phase II with a power low of index $\sim 2.34$, the enhancement of the averaged TDE rates for our fiducial values would become $\sim 100$, which is roughly consistent with that given by \citet{liu13}. For M32 like galaxy with central SMBH mass $M_{\rm BH} \simeq 3\times 10^6 M_\odot $, the enhancement of the TDE rates
in phase II over in Phase I is about $\sim 66$ for the averaged or $\sim 180$ for the peak TDE rates for $\gamma=2.0$. It is also roughly consistent with the results given by \citet{liu13}.

By using numerical scattering experiments, \citet{chen09,chen11} investigated the TDE rates by hard SMBHBs with extremely un-equal masses at dynamical evolution corresponding to the late half-stage of Phase II. They showed that the TDE rate by an SMBHB with mass $10^7 M_\odot$ and mass ratio $q=0.01$ can be enhanced by up-to 3-4 orders of magnitude higher than that for a single SMBH fed by two-body relaxation. This is about one to two orders of magnitude higher than the peak rates obtained from our fiducial model. One of the reasons is that they assumed $\gamma=2$ in their calculations. When a $\gamma=1.5$ is adopted for the same system, the peak TDE rates decrease to about a few times $10^{-3}$ events per year as shown in Fig.~20 of \citet{chen11} , which is only a few times higher than the peak TDE rate $\sim 1.4 \times 10^{-3}$ per year estimated by extrapolating the values of our fiducial model to $\gamma=1.5$ with the fitted power-law extrapolation. Another reason for the differences may be that the enhancement of TDE rates is less pronounced for less unequal mass binaries \citep{chen09}. Our results are based on $q=1$, which is much larger than the largest mass ratio $q=0.1$ investigated by \citet{chen11}. Anyway, our \nbody simulations indicate that the enhancement effect for equal mass binaries is still significant and can be as large as a factor of a few hundred times for SMBHB systems with $M_{\rm BH} \sim 10^7 M_\odot$ and $\gamma=2.0$.

\subsection{Extrapolation for phase III}
\label{Extra_PIII}

In Phase III, SMBHBs are hard binaries with averaged separations $r_{\rm BH} \sim 3\times 10^{-3} \rinf$ -- $3\times 10^{-2} \rinf$. As was indicated in Section~\ref{res}, it is difficult analytically to obtain a physically meaningful extrapolation relation for Phase III, because the tidal loss cone is neither empty nor full. Since $\rinf \sim 0.1$ in phase III, the separation corresponds to $3\times 10^{-4} - 3\times 10^{-3}$, which is very close to the tidal radius $\rt=5\times 10^{-4}$ in our fiducial model. That means in the late stage of phase III, stars are tidally accreted in our simulation which would be ejected by slingshot, if $\rt$ would be much smaller. As mentioned by \citet{chen08}, if we only consider unbound stars or bound stars with eccentricity close to unity, the TDE rate is proportional to the geometrical cross section between stars and SMBHs. This cross section, by approximation, is proportional to $\rt$. On the other hand, the rate of stars ejected through three body interaction is proportional to the semi-major axis of SMBHB. Thus when the $r_{\rm {BH}}$ close to $\rt$, almost all of stars interacts with SMBHB will be tidally disrupted. This effect happens in reality only much later than in our simulation, due to the much smaller $\rt$ in real systems. Due to our rather large $\rt$ we may underestimate the rate of slingshot ejections and overestimate the rate of TD. So, one should be aware that in phase III our extrapolation method is more uncertain than in phase I and II.

Here we fit the \nbody simulation results with power law and estimate the TDE rates for real system of galaxy mergers by extrapolating the simulation results with fitted power law relation. The relation between TD accretion rates $\dot{M}$ and tidal radius $\rt$ is given in the bottom left panel of Fig.~\ref{fig:fit}. It shows that the simulation results for different $\rt$ can be well fitted by a presumed power-law of $\dot{M}= 1.90\times 10^{-2} r_{\rm t}^{0.8634}$, which is not seriously different from the linear relation predicted by simple cross section analysis. We also fit the numerical results $\dot{M}$ as a function of particle number $N$ with a power law and obtain $\dot{M} \simeq 1.03\times 10^{-4} N^{-0.10}$. The weak dependence of TD accretion rates on particle number $N$ implies that the tidal loss cone in Phase III is indeed not empty but close to be full.

We estimated the TDE rates for a few typical real systems of galaxy mergers by extrapolating the numerical results with fitted power-law relations. For our fiducial model, the separation of SMBHB at Phase III ranges from about 0.1 pc -- 1 pc and the TD accretion rate is $\dot{M} \sim 7\times10^{-5} \accrate$, which is about an order of magnitude higher than that of two separated single SMBHs fed by two-body relaxation in Phase I. This is inconsistent with the results of hard SMBHB given in the literature \citep{chen08}. \citet{chen08} showed that because of the three-body slingshot effect, the TDE rate in hard SMBHB system is more than an order of magnitude smaller than that in a single SMBH system if the loss cone refilling is dominated by two-body interaction. This inconsistency can be understood as follows. First, as discussed above, the extrapolation based on our \nbody simulations tends to overestimate TDE rate in phase III. Second, when an SMBHB becomes hard, the nuclear stellar system is out of equilibrium and evolving with time. That leads to significantly higher tidal loss cone refilling rates comparing to the predictions based on two-body interaction.

For a galaxy like M32 with SMBH mass $M_{\rm BH} \sim 3\times 10^6 M_\odot$, the separation of SMBHB at Phase III ranges from about 0.01 pc to 0.1 pc and the extrapolated TD accretion rate is $\sim 2 \times10^{-4}\accrate$, which is about 5 times higher than that in Phase I. While, for a galaxy like M105, the extrapolated TD accretion rates is $6\times10^{-5}\accrate$, about 12 times higher than that in Phase I. The TD accretion rate in Phase III also depends on $\gamma$. It can be well fitted by a power law of index $\sim 1.49$. This gives an estimate of TD accretion rate for a M32-like galaxy with $\gamma=2.0$, $\sim 6\times10^{-4}\accrate$, which is about 3 times higher than Phase I. Therefore, the enhancement of TD accretion rates between Phases I and III  only weakly depends on $\gamma$.

\section{Implications for Observations}
\label{obv}

As our results have shown, the TDE rate of merging galaxy can be boosted a lot. For our fiducial model, it can be boosted up to 80 times, which means the TDE detection in merging galaxies should be easier. For a dry major merger like we discussed here, the typical evolution time is $\sim 1\Gyr$ \citep{colp14}. For our extrapolated fiducial model, the period for phase II is $\sim 13\Myr$, corresponding to $\sim 2600$ TDEs. If we simply assume that the TDE rate for the rest of the time is equal to the rate in phase I, the corresponding TDEs should be $\sim 6900$. That means more than a quarter of the TDEs in a merging galaxy is prompted in phase II, which occupies only $\sim 1\%$ of the entire evolution time. However, this result is only based on major merger, which is relatively scarce in the evolution history of a galaxy. According to the results by \citet{cast14}, the major merger rate for the local universe is only a few percent galaxy$^{-1}\Gyr^{-1}$. Considering the upcoming optical transient survey, the Large Synoptic Survey Telescope (LSST)\footnote{http://www.lsst.org/lsst}, may detect several hundreds of TDEs every year \citep{wegg11}, there should be several TDEs corresponding to merging galaxy in phase II can be found. For minor mergers which are more common in the lifetime of a galaxy \citep{hopk10}, as referred above, it is hard to make predictions. In this paper we have only studied major mergers, for which we found a significant boost of their TDE.

Boosted TDE rates in phase II correspond to two close SMBHs with strong perturbation to each other. A detected TDE from one of these two SMBHs in this stage should be observed as an offset flare. But the displacement from the center of the galaxy should be of order $1-10\pc$ at maximum, which is very hard to detect in distant galactic nuclei. On the other hand, such events should occur with a high velocity relative to the local rest frame in the galaxy (could be several thousand km/s), so they might be detectable through a blue or red shift of their spectral features. This kind of observational signature is similar to a TDE from recoiling SMBH, but the latter has much more smaller event rate \citep{komo08,li12,ston12}. As suggested by \citet{liu13}, it is possible to distinguish the TDE from merging galaxies and a recoiling remnant.

\section{Summary}
\label{sum}

In this paper, we investigated the dynamical evolution of SMBHBs in gas poor systems of galaxy mergers with equal mass. By using direct N-body simulations with resolution up to two million particles on the special many-core hardware $laohu$ GPU cluster at NAOC, we self-consistently calculated the TDE rates of stars by the SMBHs and its variations with the dynamical evolution of SMBHBs. The typical separation of SMBHs in galaxy mergers ranges from $r_{\rm BH} \sim 200 \rinf$ to $\sim 2\times 10^{-3} \rinf$. We found that the evolution of TDE rates by SMBHBs in galaxy mergers can be divided into three stages, which tightly correlates with the dynamical evolution of SMBHBs. In Phase I, the typical separation of two SMBHs is $r_{\rm BH} \sim 200 \rinf$ to $\sim 20 \rinf$ and the dynamical evolution of two SMBHs are dominated by the dynamical friction.  At this stage, the tidal loss cone is empty and its refilling is dominated by two-body relaxation. The TDE rate by each SMBH is about that estimated for single SMBH feeding by two-body relaxation and strongly depends on the mass of central SMBHs and the stellar density profile index $\gamma$. In Phase II, two SMBHs rapidly evolve from two strongly interacting but unbound nuclear systems to self-bound and finally hard SMBHB system, with typical separation $r_{\rm BH} \sim 20 \rinf$ to $\sim 3\times 10^{-2} \rinf$. At Phase II, the tidal loss cone is full because of the triaxial stellar distribution and the strong perturbations to the star excited by the companion. The TDE rates of SMBHBs can be enhanced by up to two order of magnitudes relative to the TDE rates in Phase I. The enhancement effect depends on the mass of central SMBH but only weakly on the stellar distribution parameter $\gamma$.

Our conclusions about Phases I and II are generally consistent with the results given in the literatures \citep{ivan05,chen09,chen11,liu13}. However, our results for Phase III is inconsistent with the results given for hard SMBHBs by \citet{chen08}. In Phase III, SMBHBs are hard and evolve slowly from a separation about $3\times 10^{-2} \rinf$ to $2\times 10^{-3} \rinf$. During this stages, the tidal loss cone is neither full nor completely empty and the loss cone feeding depending weakly on the particle number. The averaged TDE rate during Phase III is about or weakly enhanced by a factor of a few relative to the TDE rate for Phase I. This is in contrast with the conclusions given by \citet{chen08}, in which the TDE rates for hard SMBHBs are found to be by about an order of magnitude smaller than that estimated for single SMBH feeding by two-body relaxation in spherical isotropic stellar systems.

We conclude that the TDE rates in galaxy major mergers are strongly enhanced by up to two orders of magnitude relative to the estimation of single SMBH feeding by two-body relaxation, which is consistent with the high detection rate of TDEs in E+A galaxies (Arcavi et al. 2014), a type of galaxies which is probably the post-mergers of galaxies \citep{zab96,got05,ston16}. For minor mergers which are more common compare to major mergers, it is hard to make predictions without detailed numerical simulations. This issue is out of the scope of this article.

\acknowledgments

We are grateful to Xian Chen for his help on comparing our results with \citet{liu13}, and Shiyan Zhong for helpful discussions. This work is supported by the National Natural Science Foundation of China (NSFC11303039, NSFC11473003), and the Strategic Priority Research Program "The Emergence of Cosmological Structures" of the Chinese Academy of Sciences (CAS, Grant No. XDB09000000). We (SL, PB, RS) acknowledge support by CAS through the Silk Road Project at National Astronomical Observatories (NAOC) with the ¡°Qianren¡± special foreign experts program of China, and the support by Key Laboratory of Computational Astrophysics. PB was partially supported by the Sonderforschungsbereich SFB 881 "The Milky Way System" (subproject Z2) of the German Research Foundation (DFG), and by the Volkswagen Foundation under the Trilateral Partnerships grant No. 90411. PB acknowledges the special support by the CAS President's International Fellowship for Visiting Scientists (PIFI) program during his stay in NAOC, CAS and also the special support by the NASU under the Main Astronomical Observatory GRID/GPU "golowood" computing cluster project. The computations have been done on the Laohu supercomputer at the Center of Information and Computing at NAOC, CAS, funded by Ministry of Finance of People's Republic of China under the grant $ZDYZ2008-2$. Many thanks are due to the referee for valuable comments.

\end{document}